\documentclass[twocolumn,showpacs,preprintnumbers,amsmath,amssymb
 aps,
 prb,
 lengthcheck,%
]{revtex4-1}
\usepackage{amssymb}
\usepackage{graphicx}
\usepackage{dcolumn}
\usepackage{bm}
\usepackage{hyperref}
\usepackage{makeidx}

\usepackage[english]{babel} 
\usepackage[latin1]{inputenc}
\usepackage{color}
\usepackage{graphicx}
\usepackage{amsmath}

\usepackage{latexsym}
\usepackage{amsthm}

\usepackage{bbm}
\usepackage{float}

\makeindex
\def\>{\right\rangle}
\def\<{\left\langle}
\def\be{\begin{equation}}
\def\ee{\end{equation}}
\def\ba{\begin{array}{l}}
\def\ea{\end{array}}
\def\beq{\begin{eqnarray}}
\def\eeq{\end{eqnarray}}

\begin{document}

 
\title{Cooper pair splitting in a nanoSQUID geometry at high transparency}

\author{R. Jacquet,$^1$ J. Rech,$^1$ T. Jonckheere,$^1$ A. Zazunov,$^2$ and T. Martin$^1$}
\affiliation{$^1$ Aix-Marseille Universit\'e, Universit\'e de Toulon, CPT, UMR 7332, France}
\affiliation{$^2$ Institut f\"ur theoretische physik, Heinrich Heine Universit\"at, D-40225 D\"usseldorf, Germany}
\date{\today}

\begin{abstract}
We describe a Josephson device composed of two superconductors separated by two interacting quantum dots in parallel, as a probe for Cooper pair splitting. In addition to sequential tunneling of electrons through each dot, an additional transport channel exists in this system: crossed Andreev reflection, where a Cooper pair from the source is split between the two dots and recombined in the drain superconductor. Unlike non-equilibrium scenarios for Cooper pair splitting which involves superconducting/normal metal ``forks'', our proposal relies on an Aharonov-Bohm measurement of the DC Josephson current when a flux is inserted between the two dots. We provide a path integral approach to treat arbitrary transparencies, and we explore all contributions for the individual phases ($0$ or $\pi$) of the quantum dots. We propose a definition of the Cooper pair splitting efficiency for arbitrary transparencies, which allows us to find the phase associations which favor the crossed Andreev process. Possible applications to experiments using nanowires as quantum dots are discussed. 
\end{abstract}

\maketitle
\section{Introduction}

In superconductors, Cooper pairs provide a natural source of entanglement\cite{epr} in both spin and momentum space. If for some reason the constituent electrons of this pair are separated and are allowed to propagate in two different metallic leads or nanodevices such as quantum dots, one expects that the entanglement is preserved because the tunneling processes are spin preserving. This process is called cross Andreev reflection (CAR), and has been the focus of both theoretical and experimental investigations in the last two decades. The initial manifestation of CAR  was proposed theoretically for nonlocal current,\cite{byers_flatte,deutscher_feinberg,melin_current,feinberg_dirty,sauret_entangler,zaikin} as well as for non-equilibrium noise cross-correlations (current-current fluctuations between the superconductor and the two outgoing devices).\cite{anantram_datta,martin_pla,torres_martin,lesovik_martin_blatter,chevallier_splitter,rech_splitter,bena,bouchiat} Indeed, the scattering formalism of electron and hole transport showed that noise cross-correlations could be positive. Strictly speaking the positive cross-correlation signal does not constitute a rigorous proof of electrons entanglement originating from superconductors, but it certainly provides evidence of Cooper pair splitting. Alternative approaches exploiting electron energy filtering using Coulomb blockade confirmed at the same time the nonlocal spin singlet nature of electron pairs in opposite quantum dots placed in the vicinity of a superconductor, exploiting T matrix calculations. \cite{recher_sukhorukov_loss,sauret_entangler} 

The possibility of generating nonlocal entangled pairs of electrons in condensed matter settings bears fundamental applications in the context of quantum information theory. Tests of quantum entanglement followed these works, based on the idea that Bell inequality violation measurements could be implemented via noise cross-correlations with the so-called superconducting Cooper pair splitter. \cite{chtchelcatchev,sauret_bell}
Moreover, these ideas have been applied to the paradigm of quantum teleportation, \cite{sauret_epjb,long_telep} as well as in other applications.\cite{bennet,ursin}   

On the experimental side, attention has mainly focused on nonlocal current measurement on the Cooper beam splitter,\cite{beckmann} a device where typically the superconducting source of electrons is connected to two leads, sometimes via embedded quantum dots.\cite{hofstetter_nature,herrmann,schindele,hofstetter_prl107} Under specific gate voltages imposed on such dots, it is possible to trigger electronic transport in the two outgoing conductors. Only a single experiment managed to measure positive noise cross-correlations when the source of electrons was rendered superconducting.\cite{heiblum}   

The main challenge with these proposals resides in the fact that they rely on non-equilibrium measurements. Strictly speaking, these measurements, although to be saluted, constitute only an indirect evidence of Cooper pair splitting, while noise cross correlation measurements represent a considerable ordeal due to the poor signal to noise ratio, and no attempt has been tried so far to reproduce them.

\begin{figure}
\centering
\includegraphics[scale=0.15]{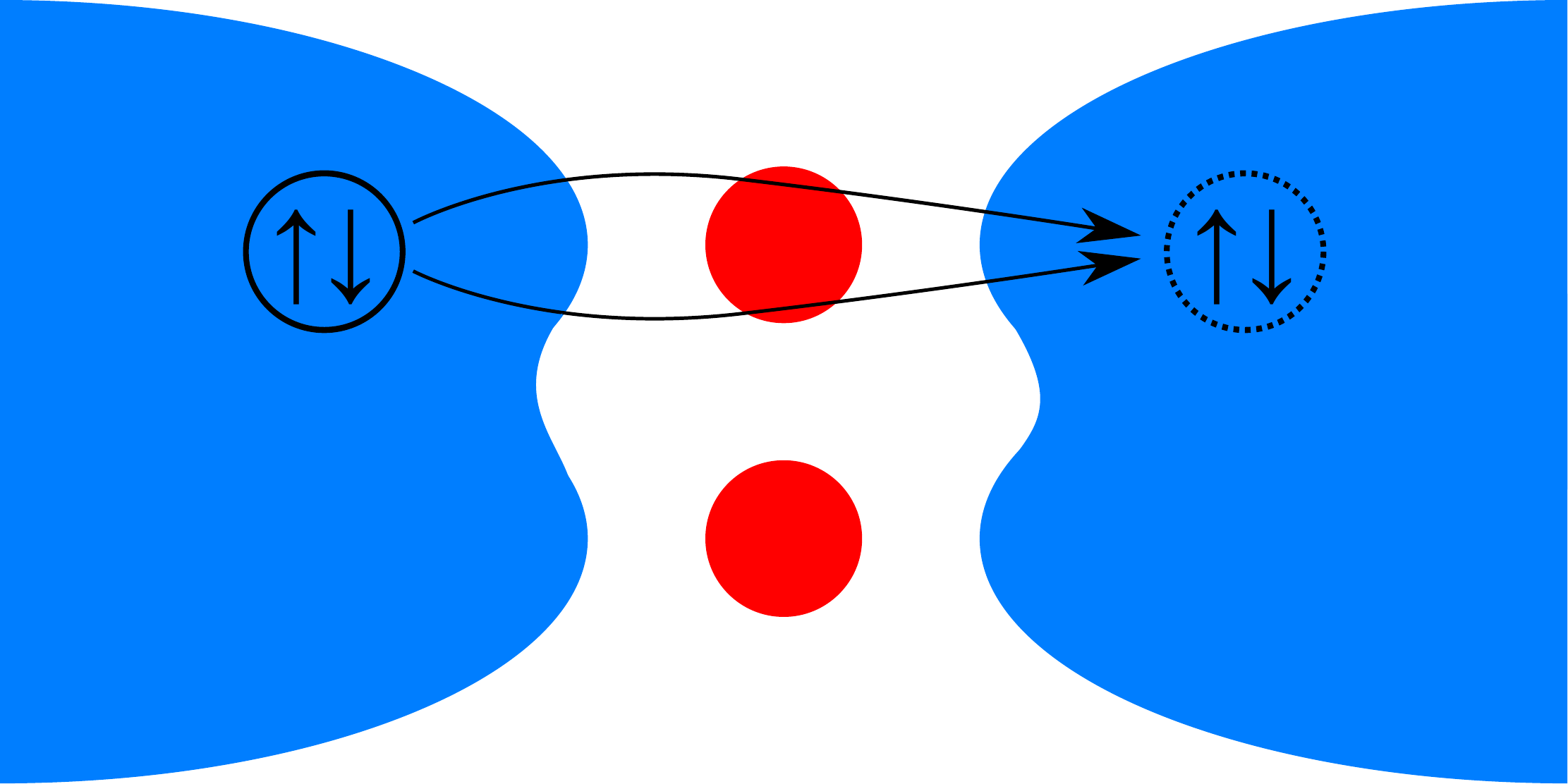}\hspace{25pt}
\includegraphics[scale=0.15]{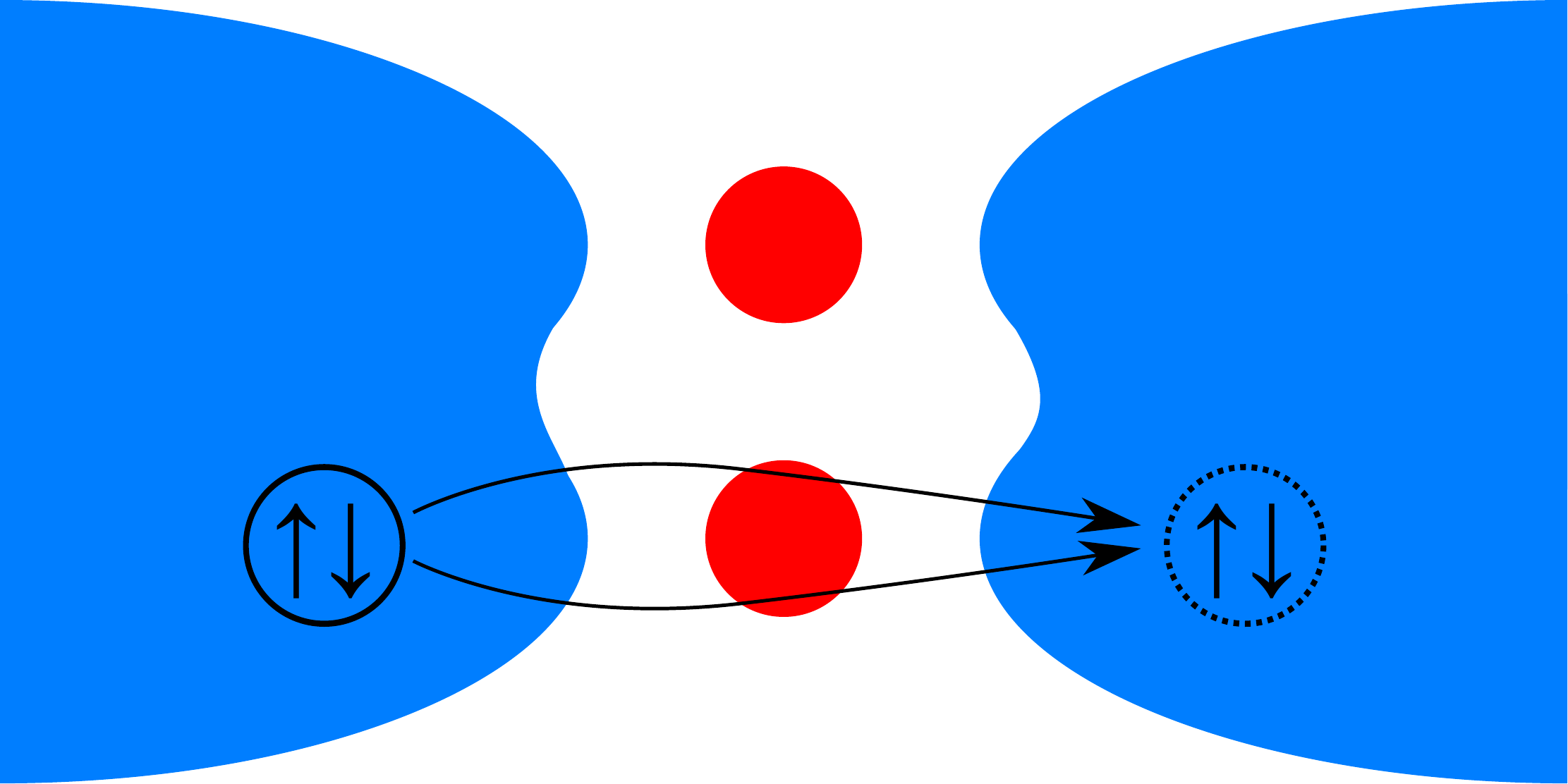}

\bigskip
\bigskip

\includegraphics[scale=0.15]{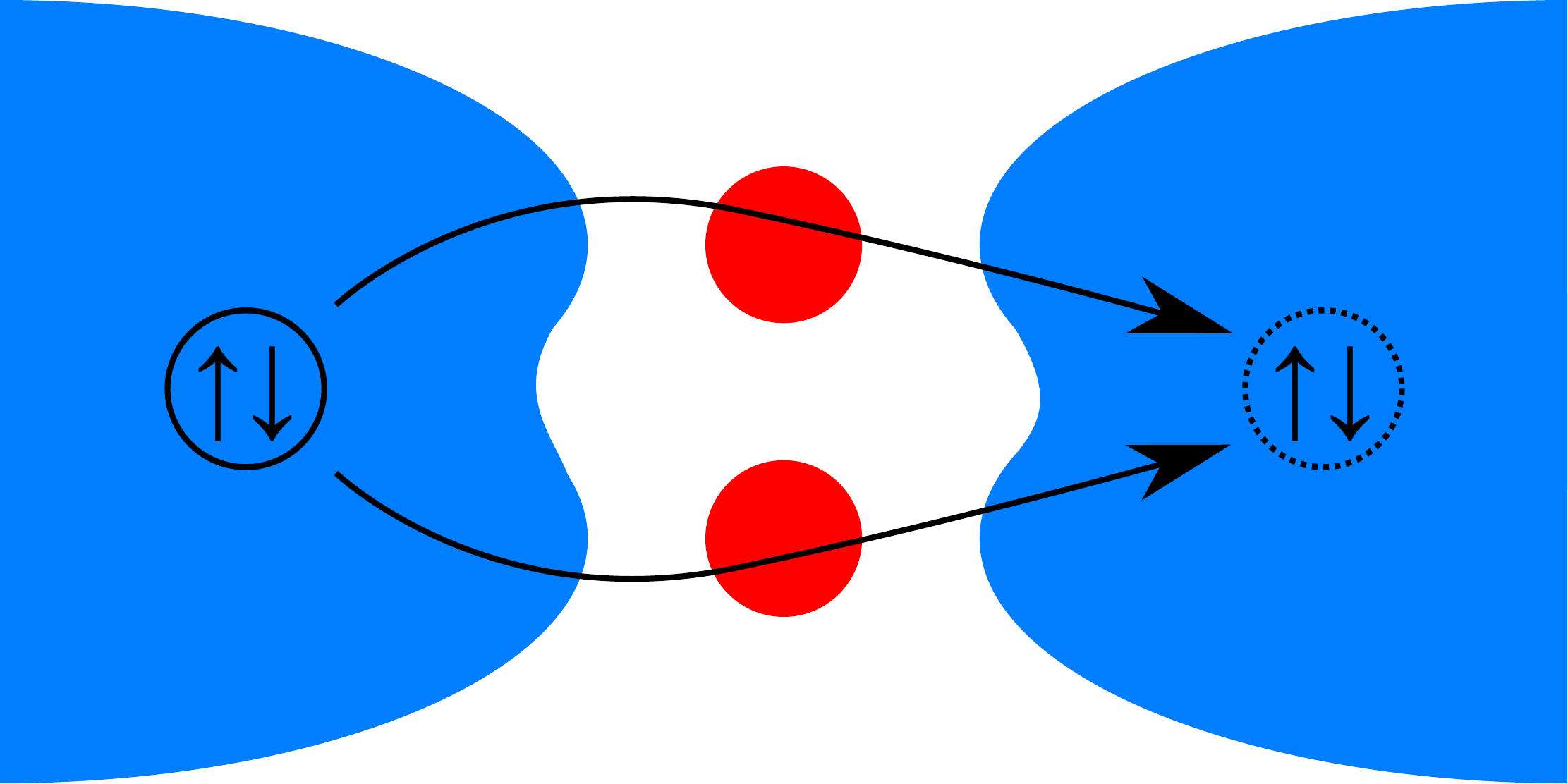}
\caption{Illustration of the three different possibilities for the transmission of a Cooper pair from
the left superconductor to the right one. The two electrons of the pair can either be both transmitted through the upper dot (top left), through the lower dot (top right), or the Cooper pair can be split with one electron transmitted through each dot (bottom).}\label{processes}
\end{figure}

A seminal theoretical work has been suggested early on to circumvent these difficulties by proposing a Josephson equilibrium current geometry to test Cooper pair splitting.\cite{choi_bruder_loss} It describes two superconductors (with applied phase difference) separated by two quantum dots placed in parallel. When a Cooper pair is transmitted from one superconductor to the other, the two electrons can either pass both through a given dot, or they can transit through different dots (cf. Fig.~\ref{processes}). This indeed realizes an Aharonov-Bohm (AB) experiment with superconductors as source and drain, driven by an applied phase difference. The critical current as a function of the AB flux should be $\pi$ periodic if electrons are not split between the two dots, and $2\pi$ periodic if Cooper pair splitting is effective. The clear originality of this proposal resides in the fact that unlike non-equilibrium noise setups, here Cooper pair splitting is uncovered using a current measurement at equilibrium albeit in a Josephson geometry. 
The calculation was performed perturbatively in the tunneling Hamiltonian, with infinite repulsion on the dots. Dot gate voltages insured that on average each dot was occupied by a single electron. A complementary study appeared a decade later with the same setup\cite{wang_hu} and perturbative results for dot levels which were assumed to be above the superconducting chemical potentials and with a finite Coulomb repulsion were also presented.      

First, so far no analysis of this superconducting Aharonov-Bohm effect has allowed to go beyond lowest order perturbation theory in the tunneling Hamiltonian. Advances in superconducting device fabrication\cite{cleuziou} seem to indicate that by burying nanowires underneath superconductors, large transmission can be achieved between the resulting quantum dot and the lead. Treating the tunnel coupling to all orders of perturbation theory thus constitutes a first motivation of our study. 

Secondly, it is established that when a quantum dot is embedded in a Josephson junction, away from the Kondo regime, the strength of the on-site repulsion, the coupling to the leads and the level position of the dot determine whether it constitutes a $0$ or a $\pi$ junction (positive or negative Josephson amplitude). This has been analyzed perturbatively,\cite{spivak_kivelson} as well as with path integral formalism, followed by a saddle point approximation.\cite{rozkhov_arovas} This latter work allows to distinguish between three phases of the Josephson junction: a) the $\pi$ phase (where the dot is singly occupied); b) the $0^{(0)}$ phase (where it is unoccupied); c) and the $0^{(2)}$ phase (with double occupation). These predictions have been verified experimentally a decade ago.\cite{van_dam} 
In the Josephson setup with two dots in parallel which is studied here, each dot can either be in the $0^{(0)}$, the $0^{(2)}$ or the $\pi$ phase. In the works of Ref. \onlinecite{choi_bruder_loss} the two electrons are both in a $\pi$ junction configuration while the work of Ref. \onlinecite{wang_hu} have also considered the case where both dots are in the $0^{(0)}$ phase. However, there is to this date no systematic or comparative study specifying which combination of the dot phases may enhance or reduce the AB signal for maximal observation, even less so for arbitrarily large transparencies.     

The two above points constitute the main motivations of the present paper. In this work, we employ the path integral formalism to model the AB setup without any restrictions on the transmission properties of the sample. Indeed, we provide results both in the tunneling and the high transparency regimes, and we propose a way to measure the efficiency of CAR processes in both cases, by analyzing the AB signal. We find that the CAR processes are optimized when the two quantum dots are in the same phase.

The paper is organized as follows.
In Sec.~\ref{model} we provide a description of the device and of the model with which we describe it, and we give an expression of the partition function in terms of Grassmann variables.  
The free energy used to derive the Josephson current of this nanoSQUID in a  non-perturbative manner is presented in Sec.~\ref{free}. In Sec.~\ref{pi_shift}, we discuss the possible occupancy states of the dots. We propose the definition of a Cooper pair splitting efficiency in Sec.~\ref{split} which can be computed for arbitrary transparencies of the studied junctions. The AB signals are calculated first in the tunneling regime in Sec.~\ref{tunnel} and then in the non-perturbative regime in Sec.~\ref{non_perturb} and all possible phase associations for the dots are considered. We discuss our results in Sec.~\ref{conclusion}.

\section{Model and partition function}
\label{model}

\begin{figure}
\centering
\includegraphics[scale=0.3]{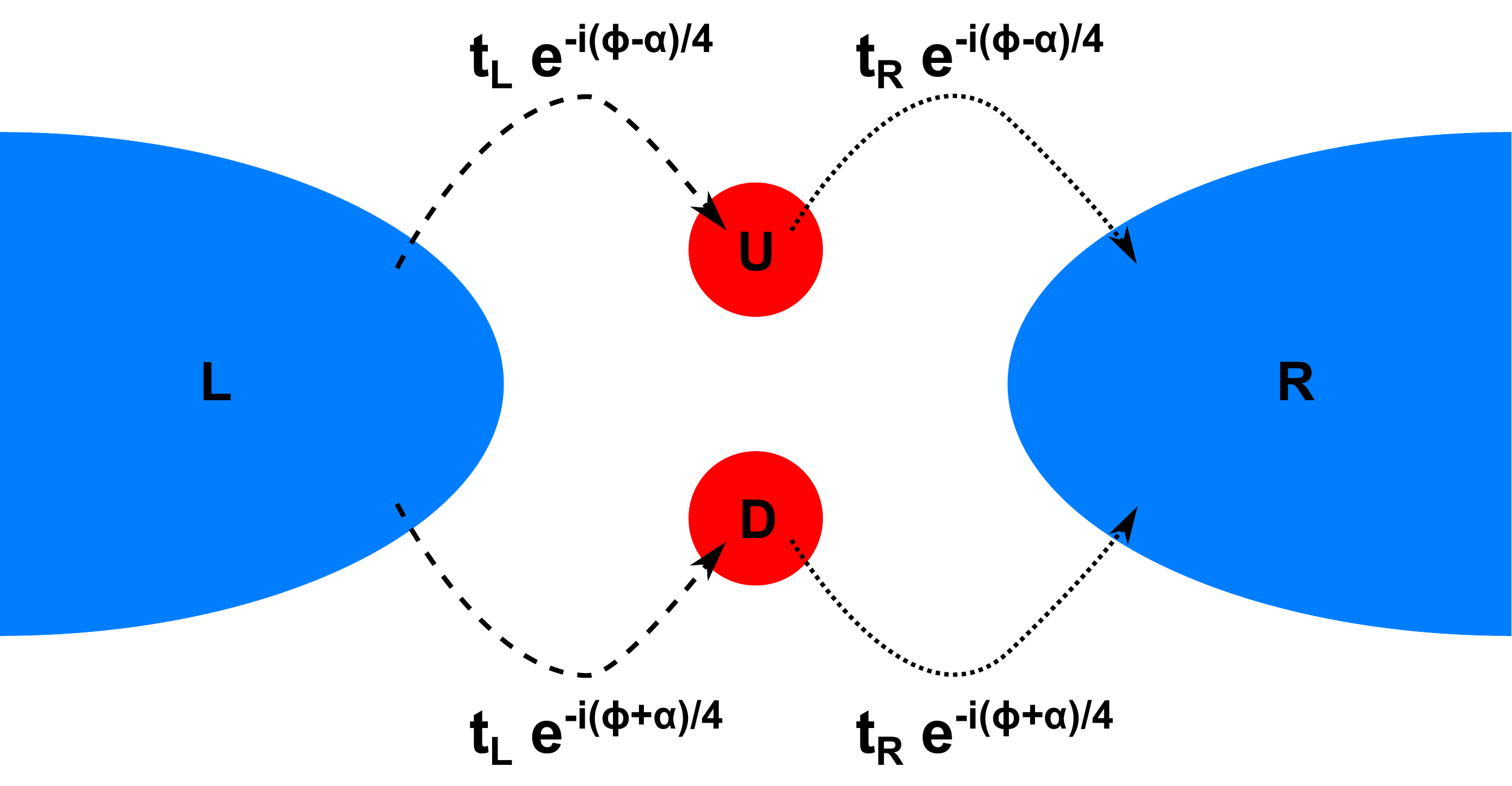}
\caption{Path dependent phase shifts.}\label{phases}
\end{figure}

The device is illustrated in Fig.~\ref{phases}. For simplicity, the two leads consist of the same superconducting material with a controllable phase difference
which can either be imposed by closing the device in a loop geometry or embedding the device in a macroscopic SQUID.\cite{choi_bruder_loss,wang_hu} Two quantum dots are placed in parallel in the nanogap between the two electrodes, and a magnetic flux (which is in principle independent from the one imposed to trigger a DC Josephson signal between the electrodes) threads the area between the two dots. Electrons can tunnel from the source electrode to the upper or lower dot, but on site Coulomb repulsion favors zero, or single occupation on the latter. In the presence of a magnetic flux between the two dots, because of the AB effect, different phase shifts are expected between the two paths that an electron can follow to reach the drain electrode. If the separation between injection points is larger than the coherence length, Cooper pairs as a whole pass either by the upper or lower path, as in the work of Ref. \onlinecite{cleuziou}. If the injection points are closer together than the superconducting coherence length however, the two electrons of the pair can travel through opposite dots, realizing a CAR process, which gives a third contribution to the Josephson current. By adjusting the quantum dot energy levels and Coulomb interaction, we expect to filter the electrons and eventually favor the CAR process. Granted, we cannot claim to directly measure the degree of entanglement of this CAR contribution (like in a Bell inequalities measurement), nevertheless, the recombination of these two electrons on the destination superconductor could not be possible if this outgoing state did not form a Cooper pair, i.e. an entangled state. 

We denote by $\hat{d}_{a\sigma}^\dag$ the creation operator for an electron with spin $\sigma=\uparrow,\downarrow$ on the quantum dot $a=U,D$ and by $\hat{\psi}_{jk\sigma}^\dag$ the creation operator for an electron with momentum $k$ and spin $\sigma=\uparrow,\downarrow$ in the superconductor $j=L,R$. It is convenient to introduce the Nambu spinors
\be
\hat{d}_a = \left(
\begin{array}{c}
\hat{d}_{a\uparrow} \\ \hat{d}^\dagger_{a\downarrow}
\end{array} \right)\quad\text{and}\quad
\hat{\psi}_{jk} = \left(
\begin{array}{c}
\hat{\psi}_{jk, \uparrow} \\
\hat{\psi}^\dagger_{j(-k), \downarrow}
\end{array} \right).\label{nambu}
\ee
$\sigma_i$ ($i=x,y,z$) are the Pauli matrices that act in Nambu space. The Hamiltonian of the double Josephson junction reads
\be
{\cal H}=\sum_{a=U,D}H_{a}+\sum_{j=L,R}H_{j}+H_t.
\ee
$H_{a}$ is the Hamiltonian of the quantum dot $a=U,D$, characterized by its energy level $\varepsilon_a$ and its on-site Coulomb repulsion $U_a$:
\begin{equation}
H_{a} = \varepsilon_a \, \sum_{\sigma = \uparrow, \downarrow} \hat{d}^\dagger_{a\sigma} \hat{d}_{a\sigma} 
+ U_a \hat{n}_{a\uparrow} \hat{n}_{a\downarrow}~,
\end{equation}
with $\hat{n}_{a\sigma}=\hat{d}^\dagger_{a\sigma} \hat{d}_{a\sigma}$ the dot occupation operator per spin.
$H_{j}$ is the Hamiltonian of the superconductor $j=L,R$, with gap $\Delta$ and chemical potential $\mu$:
\begin{equation}
H_j = \sum_k \hat{\psi}^\dagger_{jk} \left( 
\xi_k \, \sigma_z + \Delta \, \sigma_x \right) \hat{\psi}_{jk},\quad\xi_k = \frac{k^2}{2m} - \mu.
\label{ham_sc}
\end{equation}
Here $\Delta$ is real, the phase difference between electrodes has been gauged out and instead appears in the tunneling Hamiltonian. If we denote by $\mathbf{r_{ja}}$ the location of the injection point from lead $j$ to dot $a$, the tunneling Hamiltonian $H_t$ reads
\begin{equation}
H_t = \sum_{jka}\text{e}^{i\mathbf{k}.\mathbf{r_{ja}}} \, 
\hat{\psi}^\dagger_{jk} \, {\cal T}_{ja} \, \hat{d}_a + {\rm h.c.}
\label{ham_tun}
\end{equation}
The tunneling matrices involved in Eq.~\eqref{ham_tun} read
\begin{subequations}
\begin{gather}
{\cal T}_{LU}=t_L\,\sigma_z\,\text{e}^{+i\frac{\phi-\alpha}{4}\sigma_z},\quad
{\cal T}_{RU}=t_R\,\sigma_z\,\text{e}^{-i\frac{\phi-\alpha}{4}\sigma_z},\\
{\cal T}_{LD}=t_L\,\sigma_z\,\text{e}^{+i\frac{\phi+\alpha}{4}\sigma_z},\quad
{\cal T}_{RD}=t_R\,\sigma_z\,\text{e}^{-i\frac{\phi+\alpha}{4}\sigma_z}.
\end{gather}
\end{subequations}
$\phi$ is the phase difference between the superconductors while $\alpha$ is related to the magnetic flux $\Phi$ inside the SQUID loop: $\alpha=2\pi\frac{\Phi}{\Phi_0}$ where $\Phi_0=h/e$ is the flux quantum. For clarity, the phase shifts acquired by tunneling electrons are indicated in Fig.~\ref{phases}.

We employ a path integral approach in the Matsubara formalism in order to compute the partition function of the device. We introduce then the eigenvalues of the annihilation operators $\hat{\psi}_{jk\sigma}$ and $\hat{d}_{a\sigma}$ written as $\psi_{jk\sigma}$ and $d_{a\sigma}$ respectively. These are Grassmann variables and we consider also their conjugates $\overline{\psi}_{jk\sigma}$ and $\overline{d}_{a\sigma}$ as well as the collections in Nambu spinors $d_a$ and $\psi_{jk}$ defined in the same way as Eq.~\eqref{nambu}.

The partition function is given by a functional integration over paths that are $\beta$ antiperiodic:  
\be
Z=\hspace{-20pt}\int\limits_{\substack{d_a(\beta)=-d_a(0) \\ \psi_{jk}(\beta)=-\psi_{jk}(0)}}\hspace{-20pt}{\cal D}\left[\overline{d},d,\overline{\psi},\psi\right]\exp\left[-S_E\left(\overline{d},d,\overline{\psi},\psi\right)\right].
\ee
\noindent The Euclidean action $S_E$ reads
\begin{align}
S_E\left(\overline{d},d,\overline{\psi},\psi\right)&=\int\limits_0^{\beta}\text{d}\tau\,\bigg\{{\cal H}\left(\overline{d},d,\overline{\psi},\psi\right)\notag\\
&\hspace{10pt}+\sum\limits_a\overline{d}_a\partial_\tau d_a+\sum\limits_{jk}\overline{\psi}_{jk}\,\partial_\tau\psi_{jk}\bigg\}
\end{align}
where the matrix elements of the Hamiltonian can be written as
\begin{align}
\noalign{\vspace{3pt}}
{\cal H}\left(\overline{d},d,\overline{\psi},\psi\right)=&\sum_aH_{a}\left(\overline{d}_a,d_a\right)+\sum_{jk} H_{jk}\left(\overline{\psi}_{jk},\psi_{jk}\right)\notag\\
&\hspace{10pt}+\sum_{jka} H_{t,jka}\left(\overline{d}_a,d_a,\overline{\psi}_{jk},\psi_{jk}\right).
\end{align}
The expressions of $H_{jk}$ and $H_{t,jka}$ are readily obtained from Eqs.~\eqref{ham_sc}-\eqref{ham_tun} by substituting the annihilation operators $\hat{a}$ by their eigenvalues $a$ and the corresponding creation operators $a^\dag$ by the conjugate Grassmann variables $\overline{a}$. For the quantum dots, we can also find an expression in terms of Nambu spinors as follows
\be
H_{a}\left(\overline{d}_a,d_a\right)=\tilde{\varepsilon}_a+\tilde{\varepsilon}_a\,\overline{d}_a\,\sigma_z\,d_a-\frac{U_a}{2}\left(\overline{d}_ad_a\right)^2,
\ee
with $\tilde{\varepsilon}_a=\varepsilon_a+\frac{U_a}{2}$. 

\section{Free energy and Josephson current}
\label{free}

As the lead degrees of freedom are quadratic in the Hamiltonian, they can be easily integrated out. The partition function is then expressed as a functional integral over the dot Grassmann variables:
\be
Z= c_1\hspace{-10pt}\int\limits_{d_a(\beta)=-d_a(0)}\prod_a{\cal D}\left[\overline{d}_a,d_a\right]\exp\left[-S_{\text{eff}}\left(\overline{d},d\right)\right]~,
\label{partition1}
\ee
where $c_1$ is the determinant arising from the integration of the lead variables, which is independent of $\phi$ and $\alpha$. The effective action in Eq.~\eqref{partition1} reads 
\begin{widetext}
\begin{align}
S_{\text{eff}}\left(\overline{d},d\right)=&\sum_a\int\limits_0^{\beta}\text{d}\tau\left\{\tilde{\varepsilon}_a+\overline{d}_a(\tau)\left[\partial_\tau \mathbbm{1}_2+\tilde{\varepsilon}_a\,\sigma_z\right]d_a(\tau)-\frac{U_a}{2}\Big(\overline{d}_a(\tau)\,d_a(\tau)\Big)^2\right\}-\sum_{a,b}\int\limits_0^{\beta}\text{d}\tau\int\limits_0^{\beta}\text{d}\tau'\,\,\overline{d}_a(\tau)\,\Sigma^{ab}(\tau-\tau')\,d_b(\tau')\label{eff_action}
\end{align}
\end{widetext}
where the self-energy term
\begin{align}
\Sigma^{ab}(\tau)=\sum_{jk}\text{e}^{i\mathbf{k}.\left(\mathbf{r_{jb}}-\mathbf{r_{ja}}\right)}{\cal T}^\dag_{ja}G_k(\tau){\cal T}_{jb}
\end{align}
involves the Green function of the leads $G_k(\tau)$ which verifies
\be
\left[\partial_\tau \mathbbm{1}_2 +\xi_k\sigma_z+\Delta\sigma_x\right]G_k(\tau)=\delta(\tau) \mathbbm{1}_2.
\ee

The quartic terms $(\overline{d}_ad_a)^2$ in Eq.~\eqref{eff_action} prohibit an exact computation of the partition function. 
As in Ref. \onlinecite{rozkhov_arovas}, we use a Hubbard-Stratonovich transformation to treat these terms and we neglect the temporal fluctuations of the two auxiliary fields $X_U$ and $X_D$ which are introduced:
\be
\text{e}^{\frac{U_a}{2}\int\limits_0^{\beta}\text{d}\tau\left(\overline{d}_ad_a\right)^2}\approx\sqrt{\frac{\beta}{2\pi U_a}}\int_{-\infty}^{+\infty}\text{d}X_a\,\text{e}^{-\frac{\beta}{2U_a}X_a^2+X_a\int\limits_0^\beta\text{d}\tau\,\overline{d}_ad_a}.\label{quartictoX}
\ee

Because both the Green functions $G_k$ and the Nambu spinor components $d_{a\sigma}$ are $\beta$ antiperiodic, we use a Matsubara series expansion $F(\tau)=\sum_{p\in\mathbb{Z}}\text{e}^{-i\omega_p\tau}F(\omega_p)$ over the frequencies $\omega_p=\left(p+\frac{1}{2}\right)2\pi/\beta$. 
\begin{widetext}
Rather than keeping track of a cumbersome device-specific position dependence of the self-energy, we choose to introduce a phenomenological parameter $\eta$. It extrapolates between the two most relevant cases: $\eta=0$ for infinitely distant injection points (much more separated than the superconducting coherence length) and $\eta=1$ for coinciding injection points (much closer than the superconducting coherence length). The dots are now integrated out and the partition function becomes
\be
Z^\eta(\phi,\alpha)=c_1c_2\int_{-\infty}^{+\infty}\text{d}X_U\int_{-\infty}^{+\infty}\text{d}X_D\exp\left[-S^{\text{HS},\eta}_{\text{eff}}\left(X_U,X_D,\phi,\alpha\right)\right]~\label{part_func}
\ee
where $c_2$ is a fermionic determinant arising from the dot variables integration.
The effective action reads
\be
S^{\text{HS},\eta}_{\text{eff}}\left(X_U,X_D,\phi,\alpha\right)=\sum_{a}\left(\beta\,\tilde{\varepsilon}_a+\frac{\beta}{2U_a}X_a^2\right)-2\sum_{p\in\mathbb{N}}\text{ln}\Big(\beta^4\Big|\text{det}\big[{\cal M}^\eta_p\left(X_U,X_D,\phi,\alpha\right)\big]\Big|\Big),
\ee
\be
{\cal M}^\eta_p\left(X_U,X_D,\phi,\alpha\right)=\left[\begin{matrix}
-\left(i\omega_p+X_U\right)\mathbbm{1}_2+\tilde{\varepsilon}_U\,\sigma_z-{\cal A}_p(\phi-\alpha)&-{\cal B}^\eta_p(\phi,+\alpha)\\
-{\cal B}^\eta_p(\phi,-\alpha)&-\left(i\omega_p+X_D\right)\mathbbm{1}_2+\tilde{\varepsilon}_D\,\sigma_z-{\cal A}_p(\phi+\alpha)\end{matrix}\right],\label{matrixm}
\ee
\begin{align}
{\cal A}_p(\phi)&=\frac{\Gamma}{\sqrt{\Delta^2+\omega_p^2}}\Bigg[i\omega_p\,\mathbbm{1}_2-\Delta\left(\cos\frac{\phi}{2}\,\sigma_x+\gamma\sin\frac{\phi}{2}\,\sigma_y\right)\Bigg],\\
{\cal B}^\eta_p(\phi,\alpha)&=\eta\frac{\Gamma}{\sqrt{\Delta^2+\omega_p^2}}\Bigg[i\omega_p\left(\cos\frac{\alpha}{2}\,\mathbbm{1}_2+i\gamma\sin\frac{\alpha}{2}\,\sigma_z\right)-\Delta\left(\cos\frac{\phi}{2}\,\sigma_x+\gamma\sin\frac{\phi}{2}\,\sigma_y\right)\Bigg].\label{outdiago}
\end{align}
The decay rate is defined as $\Gamma=\pi\,\nu(0)(t_L^2+t_R^2)$ where $\nu(0)$ is the density of states of the leads at the Fermi level. The contact asymmetry is given by $\gamma=(t_L^2-t_R^2)/(t_L^2+t_R^2)$. 
\end{widetext}

The CAR process is due to the off-diagonal terms ${\cal B}_p$ of the matrices ${\cal M}_p$ the determinants of which we have to compute as a result of the Gaussian integrals over quantum dot degrees of freedom. In practice, these off-diagonal terms depend on the separation between injection points $R\equiv|\mathbf{r_{jU}}-\mathbf{r_{jD}}|$. They have an exponential decay on the scale of the superconducting coherence length. With the microscopic tunneling Hamiltonian formulation of Eq.~\eqref{ham_tun}, they also bear fast oscillations on the Fermi wavelength, and possibly power law decay depending on the dimensionality [$\eta(R)=(\sin k_FR)/(k_FR)$ for 3D superconductors \cite{choi_bruder_loss,popoff_these} and no power law decay for quasi-one dimensional superconductors \cite{rech_quartets}]. Existing CAR experiments\cite{hofstetter_nature,herrmann,schindele,hofstetter_prl107} based on nanowire quantum dots embedded in the superconducting leads are typically performed for an injection separation which is less than the superconducting coherence length. Yet these experiments find no evidence of either Fermi wavelength oscillations or power law decay at all: the nonlocal signal is strong and can be optimized by tuning the dot gate voltages. This can be attributed to the proximity effect from the bulk superconductors, which acts on the nanowires used in the experiments. To keep our discussion more general, and avoid such device-specific complications, in this work, we use $\eta$ as a phenomenological parameter, as in Refs. \onlinecite{wang_hu,sadovsky}. Most of the results will be displayed for the extreme values $\eta=0$ and $\eta=1$, but in order to show the evolution of the AB signals, we sometimes allow it to vary smoothly between these two values.

To evaluate the partition function of Eq.~\eqref{part_func}, we use a saddle-point method.\cite{rozkhov_arovas} The effective action $S^{\text{HS},\eta}_{\text{eff}}\left(X_U,X_D,\phi,\alpha\right)$ is computed numerically by summing over Matsubara frequencies (up to a cut-off much larger than the superconducting gap). Its minimum, located in $[X_U^\ast(\phi,\alpha),X_D^\ast(\phi,\alpha)]$, is obtained with a gradient descent method for fixed $\phi$ and $\alpha$ (the number of starting points of the algorithm depends on the symmetry of the function to be minimized). The free energy is then defined from this minimum value as $S_{\text{eff}}^{\text{HS},\eta}\left(X_U^\ast(\phi,\alpha),X_D^\ast(\phi,\alpha),\phi,\alpha\right)\equiv\beta F^\eta(\phi,\alpha)$. The current is finally obtained by differentiating the free energy with respect to the phase difference $\phi$:
\be
J^\eta(\phi,\alpha)=2\,\partial_\phi F^\eta(\phi,\alpha).
\ee
The critical current, function of the flux $\alpha$, is defined as 
\be
J_c^\eta(\alpha)=\text{max}_\phi\left|J^\eta(\phi,\alpha)\right|.
\ee
For $\eta=0$, it is $\pi$ periodic. Indeed, in this case, there are no CAR processes and for a magnetic flux $\alpha$, the current characteristic (current as a function of the phase difference $\phi$) through the double junction is simply the sum of the current characteristics across the independent single junctions, one being shifted by $-\alpha$, the other one by $+\alpha$.  Shifting $\alpha$ by $\pi$ 
results in adding a phase shift $-\pi$ for one of the two junctions [${\cal A}_p(\phi-\alpha)\to{\cal A}_p(\phi-\alpha-\pi)$ in the top left block of Eq.~\eqref{matrixm}] and a phase shift $+\pi$ for the other one [${\cal A}_p(\phi+\alpha)\to{\cal A}_p(\phi+\alpha+\pi)$ in the bottom right block of Eq.~\eqref{matrixm}]. Using the $2\pi$ periodicity in $\phi$ of the current through a single junction,
and taking the maximum on $\phi$ shows immediately that the critical current in unchanged when 
$\alpha \to \alpha + \pi$. This is
in  contrast to the case with $\eta=1$, \textit{i.e.} in the presence of CAR processes, where
the off-diagonal terms of Eq.~\eqref{matrixm} implies that only the $2\pi$ periodicity of the critical current can be observed.

\section{$0-${\Large $\pi$} transition in a single Josephson junction}\label{pi_shift}

As a first step, let us recall some known results concerning a single dot embedded in a Josephson junction.
In this section, we summarize the properties of such a setup, and we determine under what condition the junction is in the $0^{(0)}$, the $\pi$ or the $0^{(2)}$ phase. The 0 phase is characterized by a positive Josephson current for $\phi\in[0,\pi]$. It can be further separated between a $0^{(0)}$ phase where the dot is almost empty and a $0^{(2)}$ phase where the mean occupation number of the dot is almost 2. The $\pi$ phase is associated with a negative Josephson current for $\phi\in[0,\pi]$, and corresponds to a singly occupied dot.

In perturbative calculations, the Josephson current flowing through the quantum dot is the result of the tunneling of Cooper pairs, which requires a fourth order perturbative expansion in the tunneling amplitudes between a superconductor and the quantum dot. At this order, the current can be written as $J=J_0\sin\phi$, where $\phi$ is the phase difference between the superconductors, and the sign of $J_0$ determines the 0 or $\pi$ character of the junction. Providing a single occupancy on the quantum dot ($0 < -\varepsilon \ll U$) $J_0$ can be negative\cite{spivak_kivelson} unlike the case of an empty quantum dot. 

\begin{figure}
\centering
\includegraphics[scale=0.65]{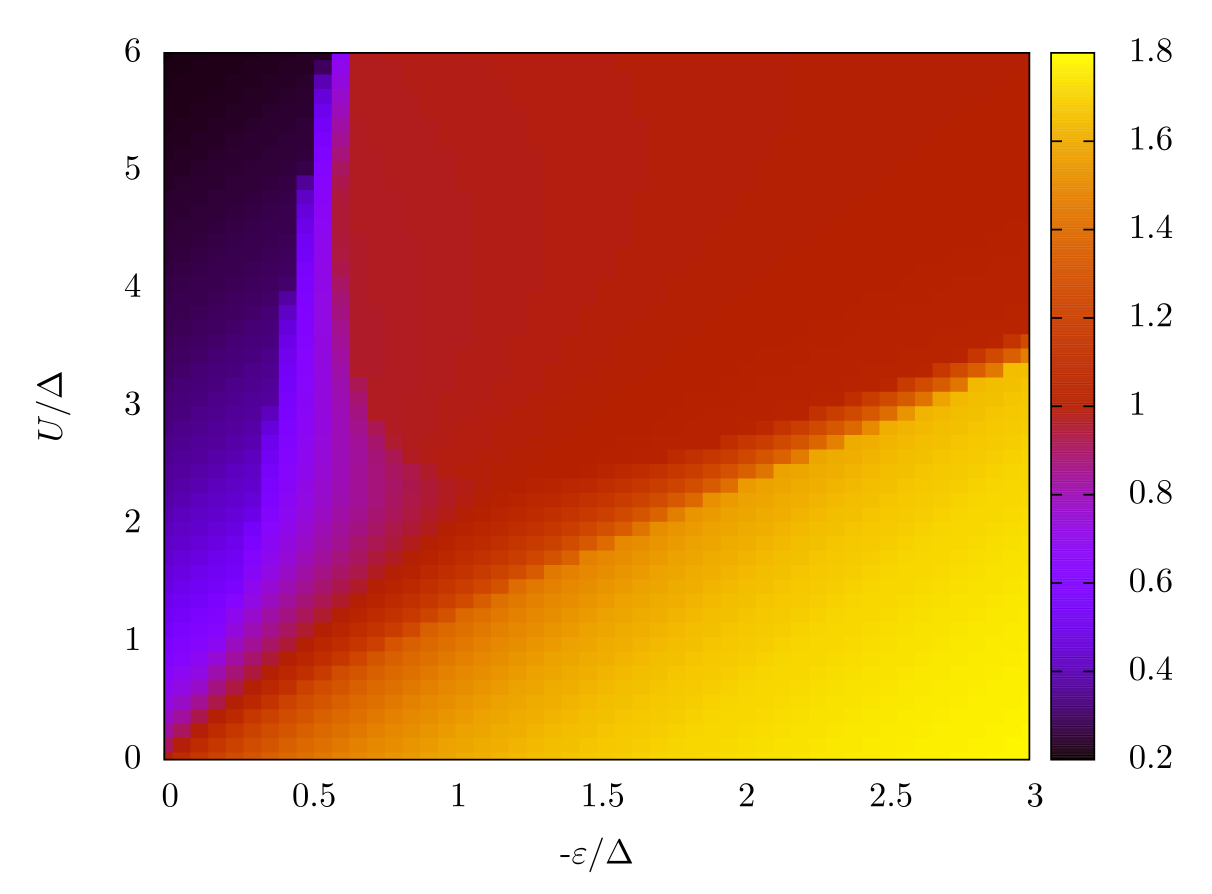}
\caption{Mean occupation number diagram of the quantum dot of a single Josephson junction for symmetric couplings ($t_L=t_R$) and for $\Gamma=\Delta$, $\beta=50\Delta^{-1}$.}\label{occupancy_Gamma}
\end{figure}

This phenomenon has been investigated numerically\cite{rozkhov_arovas} for arbitrary transmissions between the dot and the superconducting leads. For a fixed quantum dot energy level $\varepsilon<0$ and a fixed decay rate $\Gamma$, the current as a function of the phase difference $\phi$ between the superconductors undergoes a discontinuity as one tunes the Coulomb interaction $U$ across a critical value. Computing the mean occupation number on the quantum dot in the $(-\varepsilon,U)$ plane reveals the presence of all three phases, as shown in Fig.~\ref{occupancy_Gamma}, where the $\pi$ phase, which lies around the line $2\varepsilon+U=0$, separates the $0^{(0)}$ and $0^{(2)}$ phases.

\section{Splitting efficiency}\label{split}

In order to identify the optimal regime for observing signatures of CAR processes in the AB signal, we need to define a splitting efficiency. We start by discussing the case of low electron transmission, where intuition can be gained from simple perturbation theory. We then aim at comparing the results of our general approach for arbitrary transmission to those perturbative results for the AB signal.

The fourth order perturbation expansion~\cite{choi_bruder_loss,wang_hu} in the tunneling amplitudes $t_j$ allows us to write the Josephson current as three contributions, associated with the three different processes illustrated in Fig.~\ref{processes}, so that
\be
J(\phi,\alpha)=I_D\sin\left(\phi+\alpha\right)+I_U\sin\left(\phi-\alpha\right)+I_{\text{CAR}}\sin\phi
\label{curr_perturb_expr}.
\ee
Indeed, in the presence of a magnetic flux $\alpha\neq0$, an additional phase shift is acquired, depending on the path that each electron of the Cooper pair followed between the two superconducting leads.
When the Cooper pair tunnels through the $U$ (resp. $D$) quantum dot, it accumulates a phase shift $-\alpha$ (resp. $+\alpha$) in addition to the superconducting phase difference and contributes to the Josephson current with an amplitude $I_U$ (resp. $I_D$).  However, when the Cooper pair is delocalized on the two quantum dots (via the CAR process, thus contributing with an amplitude $I_\text{CAR}$), one electron gets a phase shift $+\frac{\alpha}{2}$ while the other one gets a phase shift $-\frac{\alpha}{2}$, the pair accumulating as a result no additional phase shift.

The critical current associated with the Josephson current Eq.~\eqref{curr_perturb_expr} then reads
\be
J_c(\alpha)=I_0\sqrt{1+a\,\cos\alpha+b\,\cos2\alpha}\label{crit_curr}
\ee
with $I_0=\sqrt{I_U^2+I_D^2+I_{\text{CAR}}^2}$, $I_0^2\,a=2I_{\text{CAR}}\left(I_U+I_D\right)$ and $I_0^2\,b=2I_DI_U$. For low enough values of $\Gamma$, we are in the tunneling regime and the approximation Eq.~\eqref{curr_perturb_expr} for the Josephson current is justified. 
If we are able to extract the parameters $I_U$, $I_D$ and $I_{\text{CAR}}$, e.g. from a fit of the AB signal, we can calculate the quantity
\be
r_t=\frac{I_{\text{CAR}}^2}{I_U^2+I_D^2+I_{\text{CAR}}^2}\label{spliteff}
\ee
which encodes the splitting efficiency of the double Josephson junction. It varies between 0 and 1. $r_t=0$ corresponds to a low efficiency of the CAR process while $r_t=1$ is obtained when this process of spatial delocalization of the two electrons of a Cooper pair is much more important than the tunneling processes of a whole Cooper pair through a single quantum dot. 

As already stressed out, the formalism developed in Secs.~\ref{model}-\ref{free} is valid regardless of the strength of the coupling between the superconductors and the quantum dots. However, the above definition of the splitting efficiency relies on the expression Eq.~\eqref{crit_curr} for the critical current which is no longer valid in the non-perturbative regime. There, one needs an alternative diagnosis for the detection of the CAR process. As it turns out, a relevant quantity can be extracted from the mean powers of the critical current  obtained for infinitely distant injection points ($\eta=0$) and for coinciding ones ($\eta=1$). Indeed, defining ${\cal P}_\eta=\int_0^{2\pi}\text{d}\alpha\left[J_c^\eta(\alpha)\right]^2$, we compute the quantity
\be
r=\frac{\left|{\cal P}_{\eta=1}-{\cal P}_{\eta=0}\right|}{{\cal P}_{\eta=1}}\label{spliteff2}.
\ee
This generalizes the concept of splitting efficiency to the case of arbitrary transmission, and coincides with the definition of Eq.~\eqref{spliteff} in the tunneling regime.

\section{Nanosquid in the tunneling regime}\label{tunnel}

\begin{figure}[b]
\centering
\includegraphics[scale=0.65]{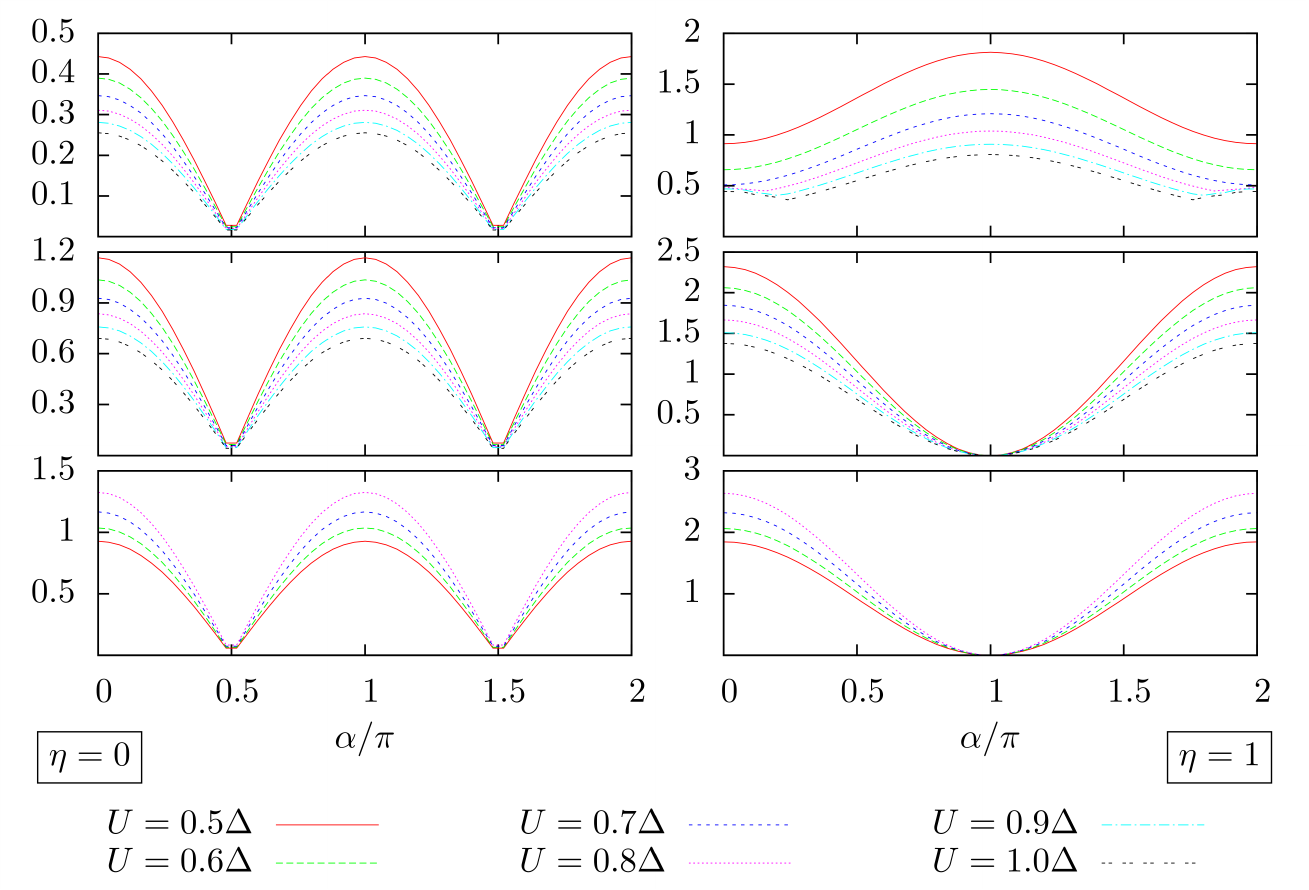}
\caption{Critical current (in units of $10^4e\Delta/\hbar$) curves for symmetric associations of dots in the tunneling regime: $\pi-\pi$ (top panel), $0^{(0)}-0^{(0)}$ (middle panel), $0^{(2)}-0^{(2)}$ (bottom panel).}\label{crit_perturb} 
\end{figure}

\begin{figure}[t]
\centering
\includegraphics[scale=0.65]{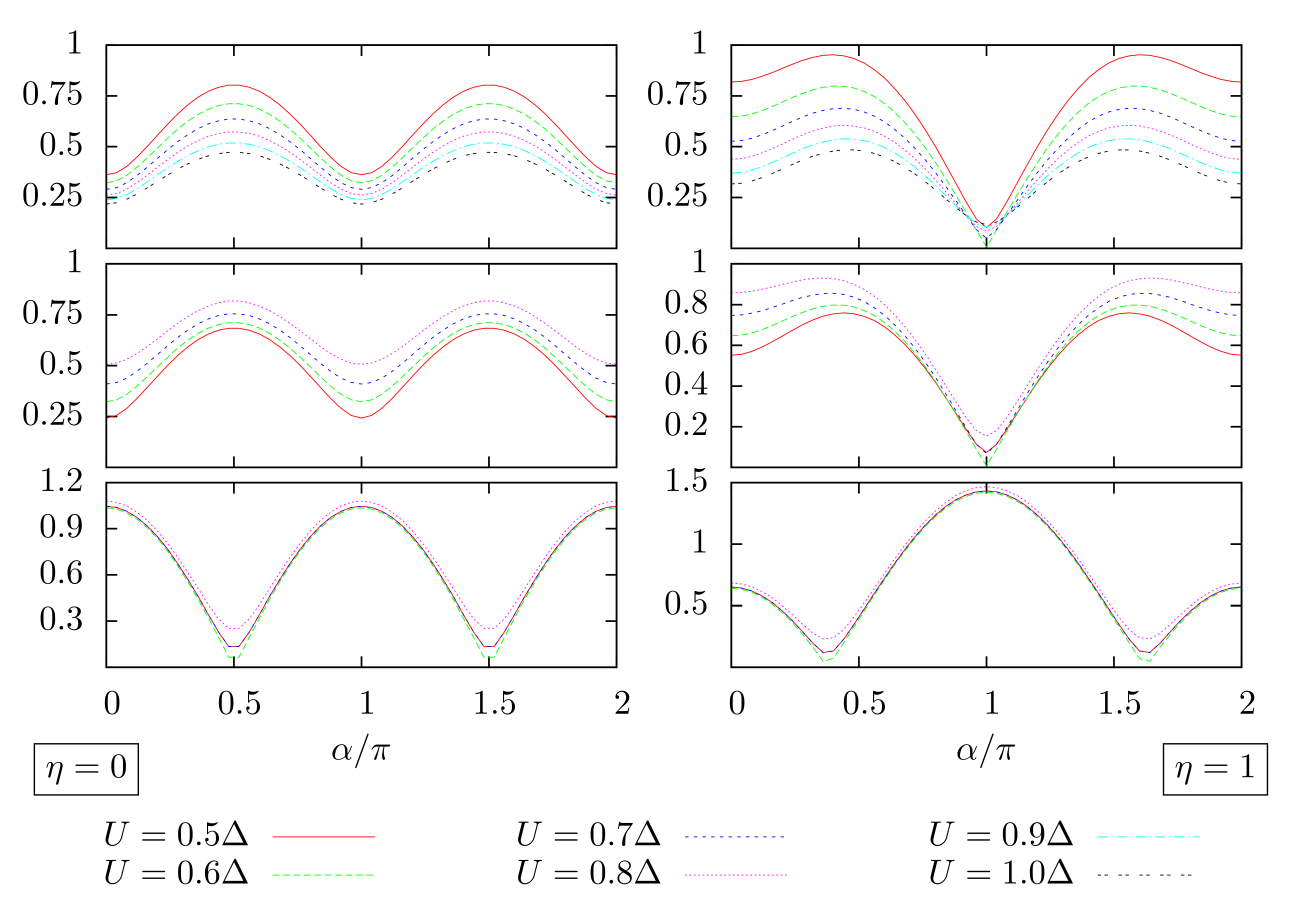}
\caption{Critical current (in units of $10^4e\Delta/\hbar$) curves for asymmetric associations of dots in the tunneling regime: $\pi-0^{(0)}$ (top panel), $\pi-0^{(2)}$ (middle panel), $0^{(0)}-0^{(2)}$ (bottom panel).}\label{crit_perturb2} 
\end{figure}

We first focus on the tunneling regime, taking a low value of the decay rate $\Gamma=0.01\Delta$.
We choose to explore all the possible combinations for the phases of the two quantum dots (see Figs.~\ref{crit_perturb}-\ref{crit_perturb2}). The results were obtained for symmetric couplings $\gamma=0$, at temperature $\beta^{-1}=0.002\Delta$. The energies of the quantum dots are: $\varepsilon=-0.3\Delta$ for the $\pi$ phase, $\varepsilon=0.3\Delta$ for the $0^{(0)}$ phase and $\varepsilon=-0.9\Delta$ for the $0^{(2)}$ phase. The Coulomb interaction $U$ is chosen to be the same for the two quantum dots, and within a specific range as staying in a given phase at fixed energy restricts the possible values for $U$.

\begin{figure}[b]
\centering
\includegraphics[scale=0.65]{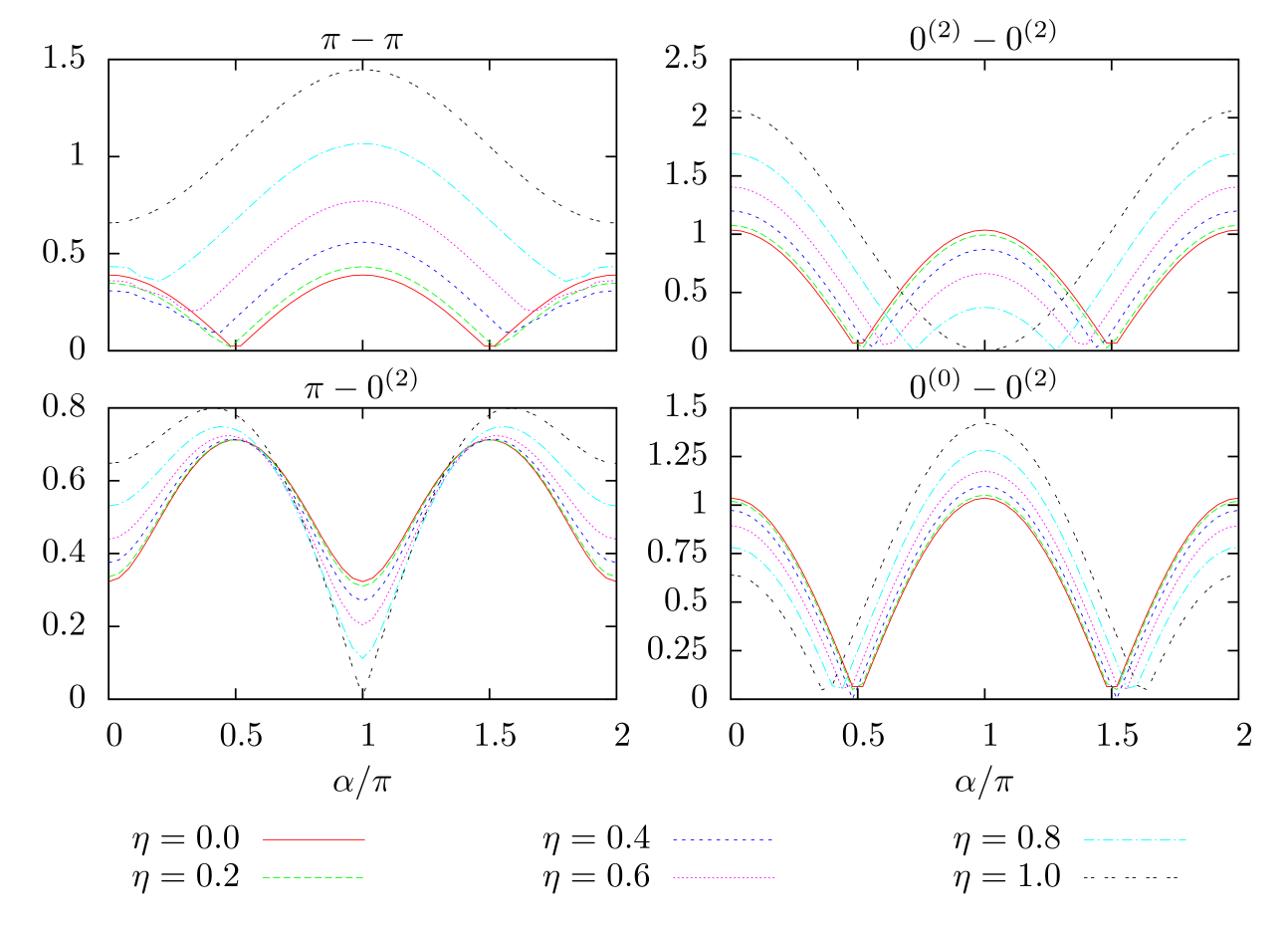}
\caption{Influence of the parameter $\eta$ on the critical current (in units of $10^4e\Delta/\hbar$) in the tunneling regime ($U=0.6\Delta$).}\label{var_eta}
\end{figure} 

\begin{figure}[b]
\centering
\includegraphics[scale=0.65]{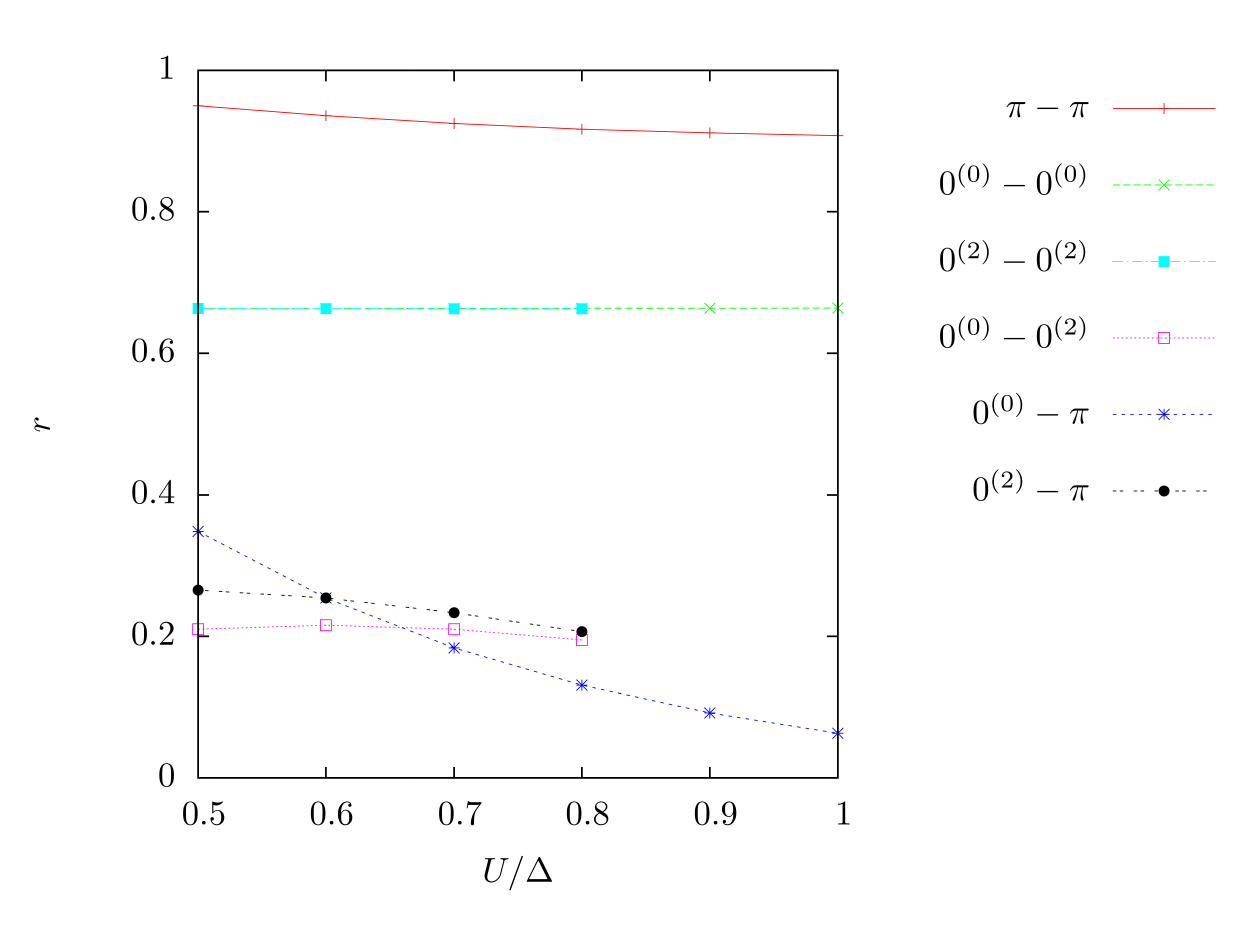}
\caption{Splitting efficiency $r$ given by Eq.~\eqref{spliteff2} extracted from the curves of Figs.~\ref{crit_perturb} and \ref{crit_perturb2}.}\label{rnum}
\end{figure}  

The particle-hole symmetry ensures that the current is invariant under the change $(\tilde{\varepsilon}_U,\tilde{\varepsilon}_D)\to(-\tilde{\varepsilon}_U,-\tilde{\varepsilon}_D)$. For the values of energy $\varepsilon$ mentioned above, this implies that the critical current is identical for $0^{(0)}-0^{(0)}$ and $0^{(2)}-0^{(2)}$ phase associations with $U=0.6\Delta$, for $\pi-0^{(0)}$ and $\pi-0^{(2)}$ phase associations again with $U=0.6\Delta$, and finally for $0^{(0)}-0^{(0)}$ at $U=0.7\Delta$ and $0^{(2)}-0^{(2)}$ at $U=0.5\Delta$.

We first consider the critical current curves of a SQUID made of two independent single Josephson junctions, \textit{i.e.} for $\eta=0$. In this particular case, $I_\text{CAR}=0$ and consequently $a=0$ in Eq.~\eqref{crit_curr}, so that the extrema of the critical current are the zeros of $\sin2\alpha$, and the total Josephson current is given by Eq.~\eqref{curr_perturb_expr} with $I_\text{CAR}=0$. 
From the results of Figs.~\ref{crit_perturb} and \ref{crit_perturb2}, it clearly appears that there is a $\pi/2$ phase shift in the critical current between the situation where the two dots are in the same phase ($0-0$ or $\pi-\pi$) and the one where they are in different phases ($\pi-0$). 
More specifically, for two quantum dots in the same phase ($I_D I_U>0$), there is no phase shift between the currents of the two single Josephson junctions for $\alpha=0$, so that the maxima are added ($|I_D+I_U|=|I_D|+|I_U|$) and, as a result, the critical current is maximal for $\alpha=0$. However, for two quantum dots in different phases ($I_D I_U<0$), the phase shift of $\pi$ between the currents of the two single Josephson junctions for $\alpha=0$ is compensated by a phase shift $\alpha=+\pi/2$ for one of the currents and a phase shift $-\alpha=-\pi/2$ for the other one ($|I_D-I_U|=|I_D|+|I_U|$) resulting in a maximum of the critical current for $\alpha=\pi/2$. Such a behavior has been observed experimentally.\cite{cleuziou}

Comparing the left panels of Figs.~\ref{crit_perturb}-\ref{crit_perturb2} to the right ones (i.e. the case $\eta=0$ to $\eta=1$), we immediately obtain evidence of the cross-talk between the two single Josephson junctions: the period of the critical current doubles. This is a signature of the emergence of the CAR process. Note that the symmetric associations (cf. Fig.~\ref{crit_perturb}) differ completely between $\eta=0$ and $\eta=1$: for the $\pi-\pi$ association, the maximum at $\alpha=0$ for $\eta=0$ becomes a minimum for $\eta=1$, and for the $0^{(0)}-0^{(0)}$ and $0^{(2)}-0^{(2)}$ associations, the maximum at $\alpha=\pi$  for $\eta=0$ becomes a zero for $\eta=1$. 
%
%
Concerning the asymmetric associations (cf. Fig.~\ref{crit_perturb2}), the critical current curves for $\eta=1$ somehow look like those obtained for $\eta=0$: the positions of the maxima and minima are mostly preserved for all phase associations, only their local or global character changes when tuning $\eta$.

Increasing $U$ results in a more pronounced filtering as the processes where the quantum dots are doubly occupied  are less favored. This explains the observed decrease in critical current for the $0^{(0)}-0^{(0)}$ and $\pi-\pi$ associations (Fig.~\ref{crit_perturb}). The opposite behavior happens when the quantum dots are doubly occupied, \textit{i.e.} for the $0^{(2)}-0^{(2)}$ association. There, increasing $U$ favors processes where the occupation of the dots is lowered, since approaching the $\pi$ transition results in the decrease of the mean occupation number on the quantum dot. Such opposite behaviors while tuning $U$ for $0^{(0)}-0^{(0)}$ and $0^{(2)}-0^{(2)}$ phase associations can be seen as a consequence of the already discussed particle-hole symmetry. Similarly, the $\pi-0^{(0)}$ and $\pi-0^{(2)}$ combinations have opposite interaction dependence (see Fig.~\ref{crit_perturb2}). Interestingly, the interaction has no noticeable effect in the case of the $0^{(0)}-0^{(2)}$ phase association.

In order to monitor the emergence of CAR processes, we investigate in Fig.~\ref{var_eta} intermediate regimes for the $\pi-\pi$, $0^{(2)}-0^{(2)}$, $\pi-0^{(2)}$ and $0^{(0)}-0^{(2)}$ phase associations by turning on $\eta$ progressively from $0$ to $1$. This is a phenomenological way to introduce more and more CAR processes, as the injection points in the superconductors are brought together, from $\eta=0$ (no CAR effect) to $\eta=1$ (maximal CAR effect). We observe a dramatic increase of the critical current in the symmetric configurations ($\pi-\pi$ and $0^{(2)}-0^{(2)}$ phase associations), while in the asymmetric configurations ($\pi-0^{(2)}$ and $0^{(0)}-0^{(2)}$ phase associations), the modification of the AB signal is noticeable, but not substantial. 

We display the splitting efficiency $r$ given by Eq.~\eqref{spliteff2} for all phase associations in Fig.~\ref{rnum}. In the studied domain for Coulomb repulsion parameter $U$, we do not observe noticeable evolutions of $r$ except for the  $0^{(0)}-\pi$ association for which we see a clear decrease. 
The symmetric associations of phases (for which the critical current differs the most between $\eta=0$ and $\eta=1$) lead to the highest values of $r$. The highest splitting efficiency is obtained for two quantum dots in the $\pi$ phase: ensuring a mean occupation number around 1 on each quantum dot favors the CAR process.
One can get some insight concerning the value of $r$ for the associations $0^{(0)}-0^{(0)}$ and $0^{(2)}-0^{(2)}$ from the perturbative calculation presented in the previous section. Indeed, in the tunneling regime the splitting efficiency $r$ matches the definition $r_t$ of Eq.~\eqref{spliteff} in terms of the amplitudes $I_D$, $I_U$ and $I_\text{CAR}$. For symmetric associations,  we must add the constraint $I_D=I_U\equiv I/2$ to the fit procedure, so that the critical current reads
\be
\frac{J_c(\alpha)}{|I|}=\left|\frac{I_{\text{CAR}}}{I}+\cos\alpha\right|
\ee
and since the critical current in symmetric associations of 0 phases vanishes for $\alpha=\pi$, we get $I_{\text{CAR}}/I=1$ and $r_t=2/3$. 

\section{Nanosquid in the high transparency regime}\label{non_perturb} 

The advantage of the formalism developed in Secs.~\ref{model}-\ref{free} relies on the possibility to address high transparency regimes of the double Josephson junction under study. The results presented in this Sec. are obtained for symmetric couplings $\gamma=0$, at temperature $\beta^{-1}=0.02\Delta$ and for a decay rate $\Gamma=2\Delta$. The energies of the quantum dots are: $\varepsilon=-4\Delta$ for the $\pi$ phase, $\varepsilon=4\Delta$ for the $0^{(0)}$ phase and $\varepsilon=-12\Delta$ for the $0^{(2)}$ phase. Particle-hole symmetry implies that the critical current is identical for $0^{(0)}-0^{(0)}$ and $0^{(2)}-0^{(2)}$ phase associations with $U=8\Delta$, for $\pi-0^{(0)}$ and $\pi-0^{(2)}$ phase associations again with $U=8\Delta$, and finally for $0^{(0)}-0^{(0)}$ at $U=9\Delta$ and $0^{(2)}-0^{(2)}$ at $U=7\Delta$.

\begin{figure}[b]
\centering
\includegraphics[scale=0.65]{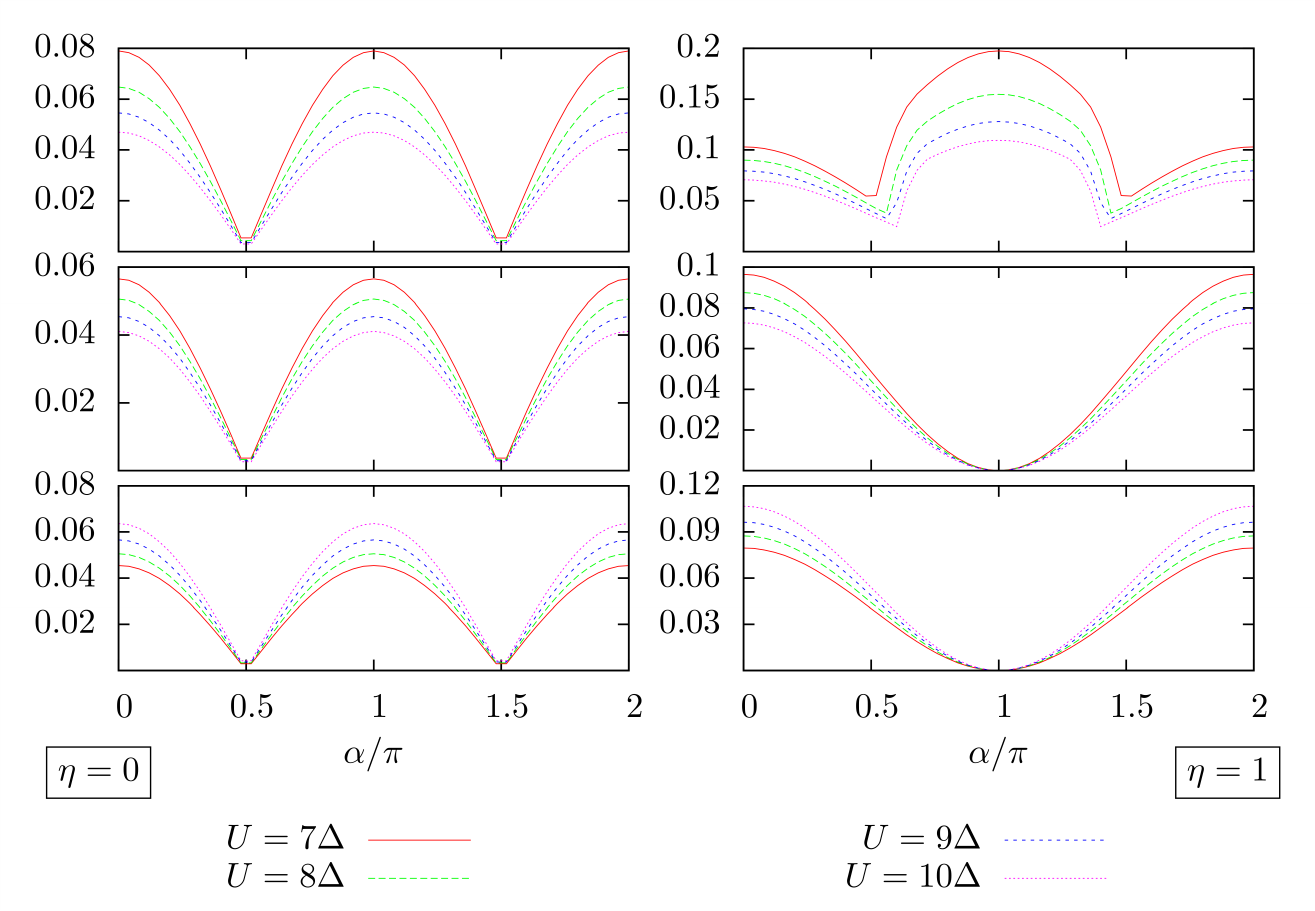}
\caption{Critical current (in units of $e\Delta/\hbar$) curves for symmetric associations of dots in the high transparency regime: $\pi-\pi$ (top panel), $0^{(0)}-0^{(0)}$ (middle panel), $0^{(2)}-0^{(2)}$ (bottom panel).}\label{crit_nonperturb}
\end{figure}

\begin{figure}[t]
\centering
\includegraphics[scale=0.65]{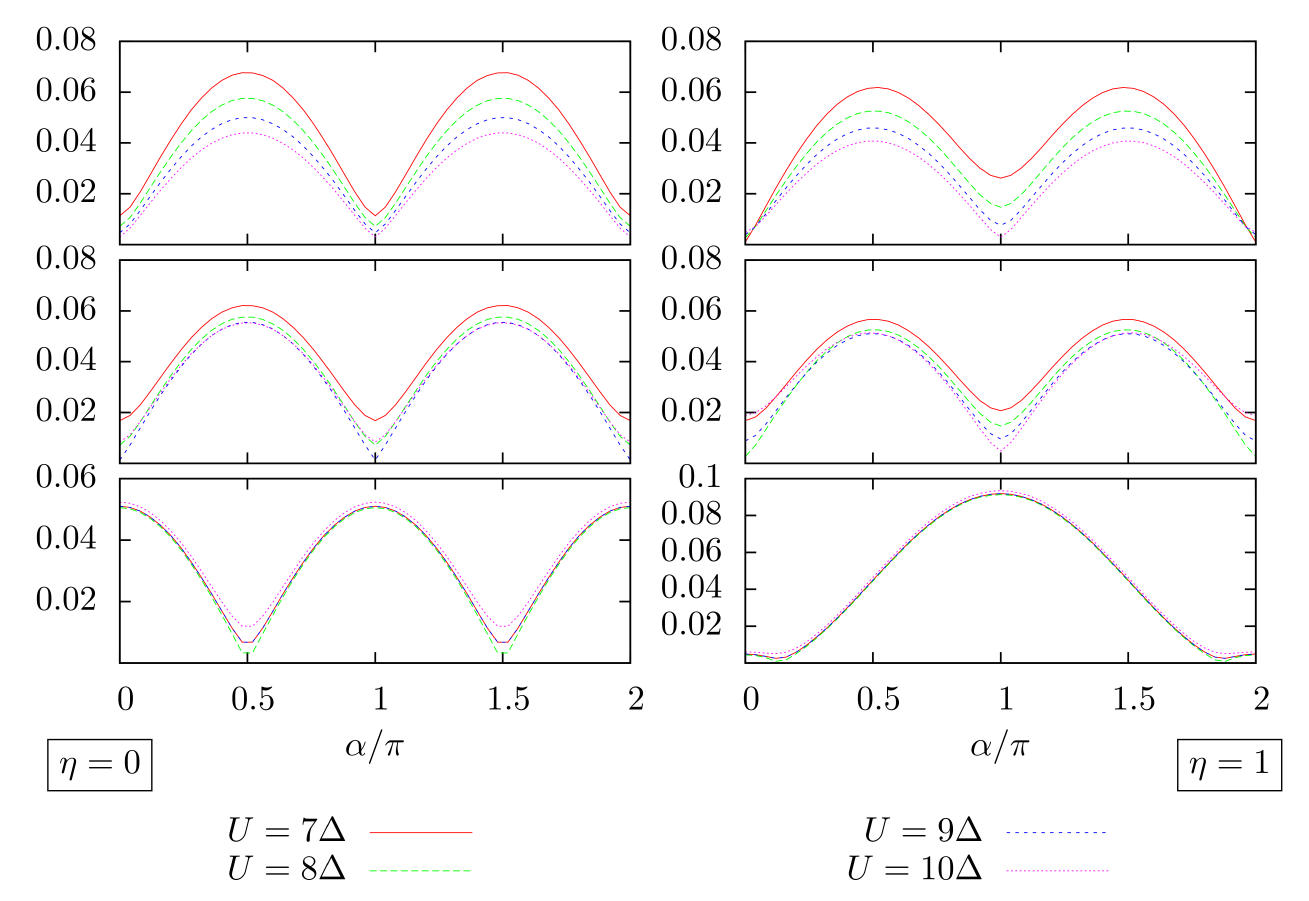}
\caption{Critical current (in units of $e\Delta/\hbar$) curves for asymmetric associations of dots in the high transparency regime: $\pi-0^{(0)}$ (top panel), $\pi-0^{(2)}$ (middle panel), $0^{(0)}-0^{(2)}$ (bottom panel).}\label{crit_nonperturb2} 
\end{figure}

The strategy is again to investigate all the possible combinations of phases for the quantum dots (which correspond to different mean occupation numbers), in order to reproduce the qualitative study of Sec.~\ref{tunnel}. Our goal is to determine what features are preserved and what has changed, and to compute the splitting efficiency $r$ defined by Eq.~\eqref{spliteff2} in order to determine which associations of phases favor nonlocal phenomena the most. The critical current curves are given in Figs.~\ref{crit_nonperturb}-\ref{crit_nonperturb2}. 

From the results of the $\pi-\pi$ association for $\eta=1$, it is clear that we are no longer in the tunneling regime as the flux dependence can not be fitted by Eq.~\eqref{crit_curr} together with the constraint that $I_D=I_U$. We can again notice, for $\eta=0$, the $\pi/2$ phase shift in the critical current between dots in the $\pi-0$ phases (top and middle panels of Fig.~\ref{crit_nonperturb2}) and dots in the $0-0$ or $\pi-\pi$ phases (Fig.~\ref{crit_nonperturb} and bottom panel of Fig.~\ref{crit_nonperturb2}).

\begin{figure}[b]
\centering
\includegraphics[scale=0.65]{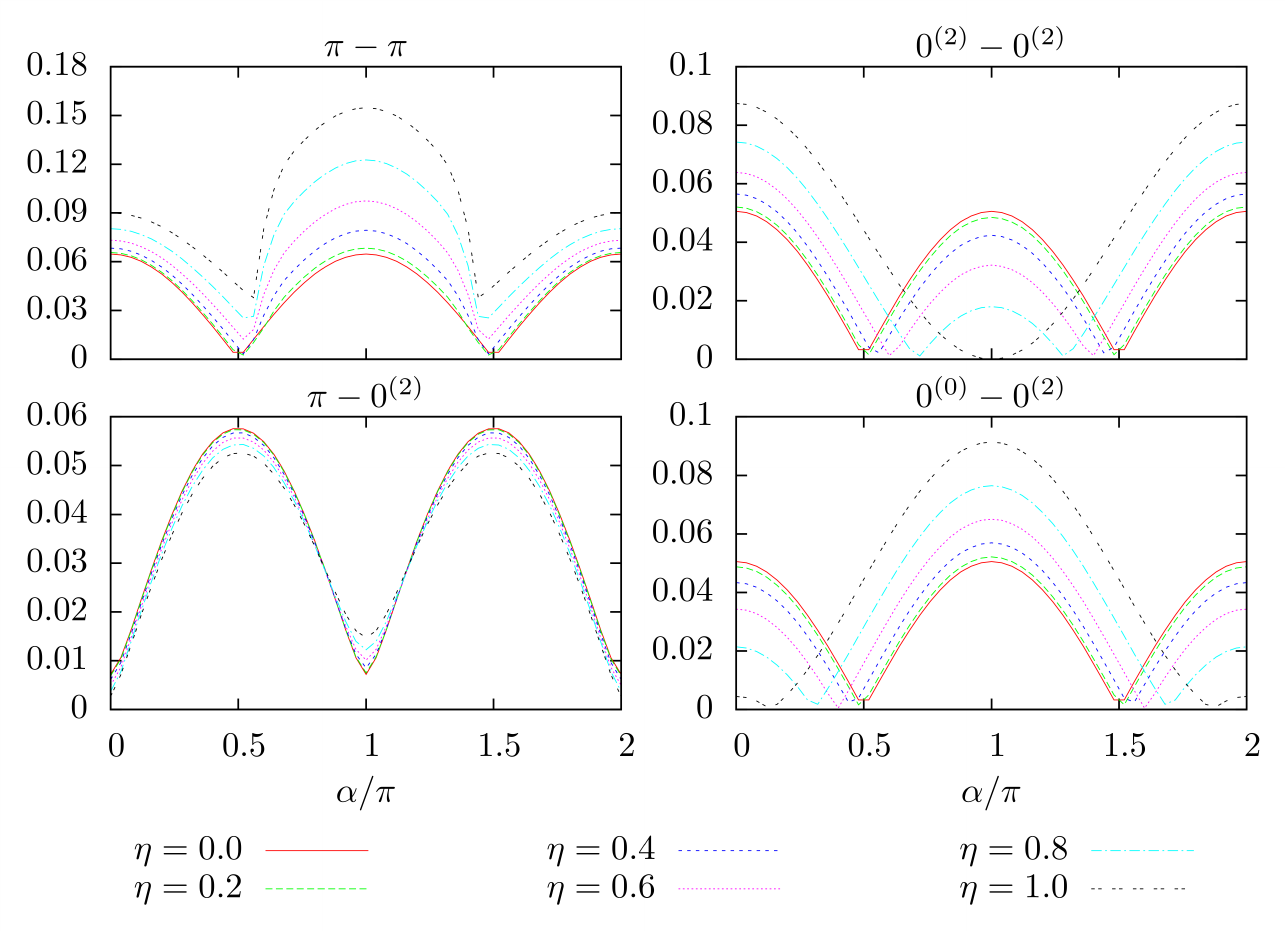}
\caption{Influence of the parameter $\eta$ on the critical current (in units of $e\Delta/\hbar$) in the high transparency regime ($U=8\Delta$).}\label{var_eta_nonperturb}
\end{figure}  

\begin{figure}[b]
\centering
\includegraphics[scale=0.65]{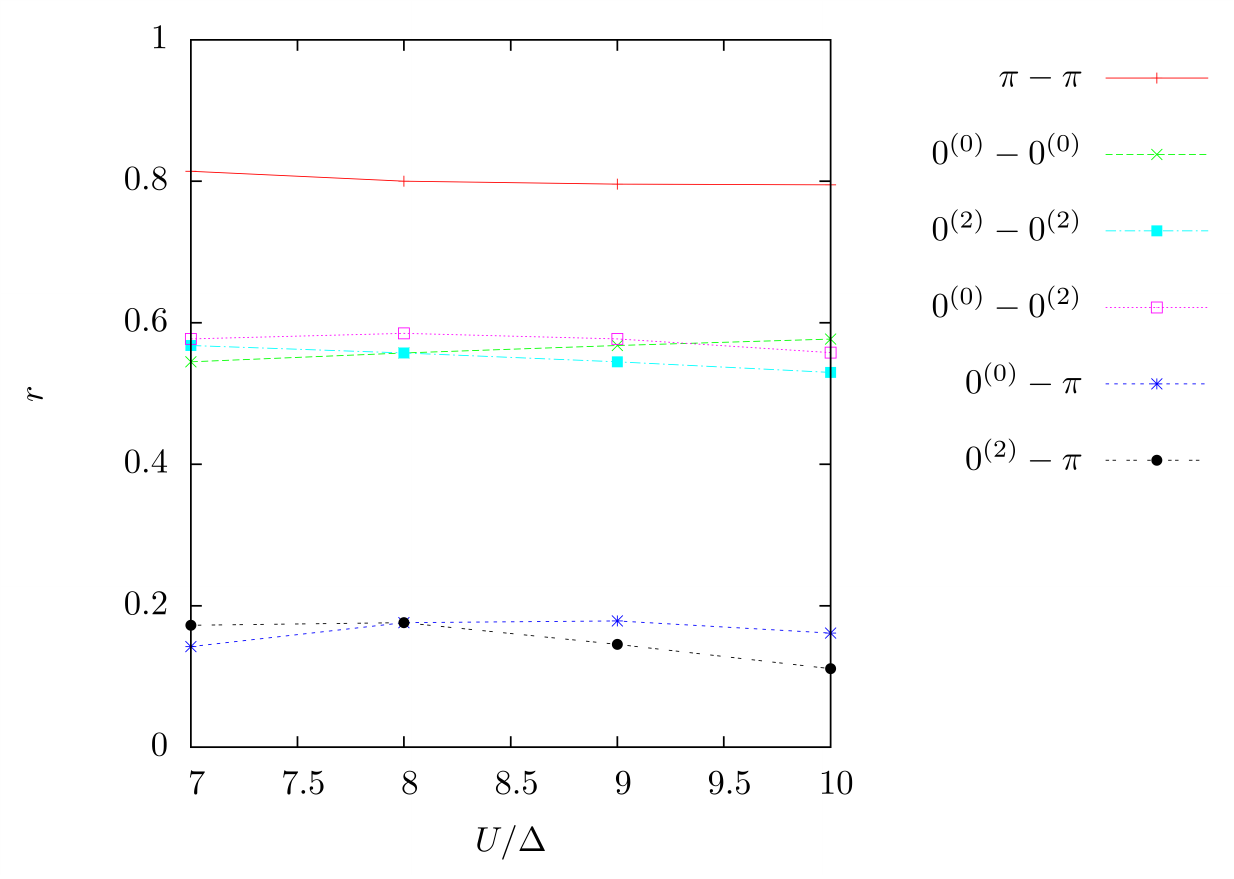}
\caption{Splitting efficiency $r$ given by Eq.~\eqref{spliteff2} extracted from the curves of Figs.~\ref{crit_nonperturb} and \ref{crit_nonperturb2}.}\label{eff_nonperturb}
\end{figure}

The critical current in Figs.~\ref{crit_nonperturb}-\ref{crit_nonperturb2} again presents a doubling of its period when switching $\eta$ from 0 to 1. This is due to the emergence of nonlocal processes where both quantum dots are involved. For the $\pi-\pi$ association, during this switching, $\alpha=0$ remains a local maximum (contrary to the tunneling regime). For $0^{(0)}-0^{(0)}$ and $0^{(2)}-0^{(2)}$ phase associations, the maximum at $\alpha=\pi$ for $\eta=0$ becomes a zero for $\eta=1$ (as in the tunneling regime). There is little influence (less than in the tunneling regime) of $\eta$ on the $\pi-0^{(2)}$ and $\pi-0^{(0)}$ associations whereas, for the $0^{(0)}-0^{(2)}$ association the maximum in $\alpha=0$ is considerably lowered (more than in the tunneling regime) from $\eta=0$ to $\eta=1$.

For symmetric associations of phases, the evolution of the critical current with increasing $U$ (Fig.~\ref{crit_nonperturb}) can be explained following the same arguments as in the tunneling regime.
While the filtering of electrons tunneling through $0^{(0)}$ or $\pi$ quantum dots is responsible for a decrease of the signal, favoring one-electron processes through a $0^{(2)}$ quantum dot results in an increase of the signal. The opposite $U$-dependence for the $0^{(0)}-0^{(0)}$ and $0^{(2)}-0^{(2)}$ phase associations is a consequence of particle-hole symmetry. While there is still no noticeable effect of $U$ on the $0^{(0)}-0^{(2)}$ association as in the tunneling regime, the $\pi-0^{(2)}$ association also shows little $U$-dependence. As a consequence, we do not observe the opposite behavior as a function of $U$ for the $\pi-0^{(0)}$ and $\pi-0^{(2)}$ associations (Fig.~\ref{crit_nonperturb2}).

We introduce progressively nonlocal effects in Fig.~\ref{var_eta_nonperturb} by switching $\eta$ from 0 to 1. 
For the symmetric associations ($\pi-\pi$ and $0^{(0)}-0^{(0)}$) as well as for the $0^{(0)}-0^{(2)}$ configuration (contrary to the tunneling regime), the AB signal is dramatically increased when CAR processes are switched on. On the contrary, the $\pi-0^{(2)}$ phase association is hardly influenced by the presence of nonlocal processes (even less than in the tunneling regime).  
We can quantify the importance of CAR processes compared to the direct tunneling through a single quantum dot by calculating the splitting efficiency $r$ given by Eq.~\eqref{spliteff2} which we display in Fig.~\ref{eff_nonperturb} for the different associations of phases. As a function of $U$, the splitting efficiencies which are found are essentially constant over the range considered. 
As it turns out, although we work with specific phases for the individual quantum dots, implying specific populations, the quantization of the electron charge on these dots is ineffective at high transparencies because they constitute ``open quantum dots'', involving large fluctuations from their average population. This is consistent with the observation that the splitting efficiency varies little with $U$. As mentioned above, there is little influence of $\eta$ on the $\pi-0^{(2)}$ and $\pi-0^{(0)}$ associations and this is why these are the lowest splitting efficiencies we found. The splitting efficiencies of the associations of 0 phases are around 0.55  and the highest splitting efficiency is still obtained for the $\pi-\pi$ association.    

\begin{figure}
\includegraphics[scale=0.65]{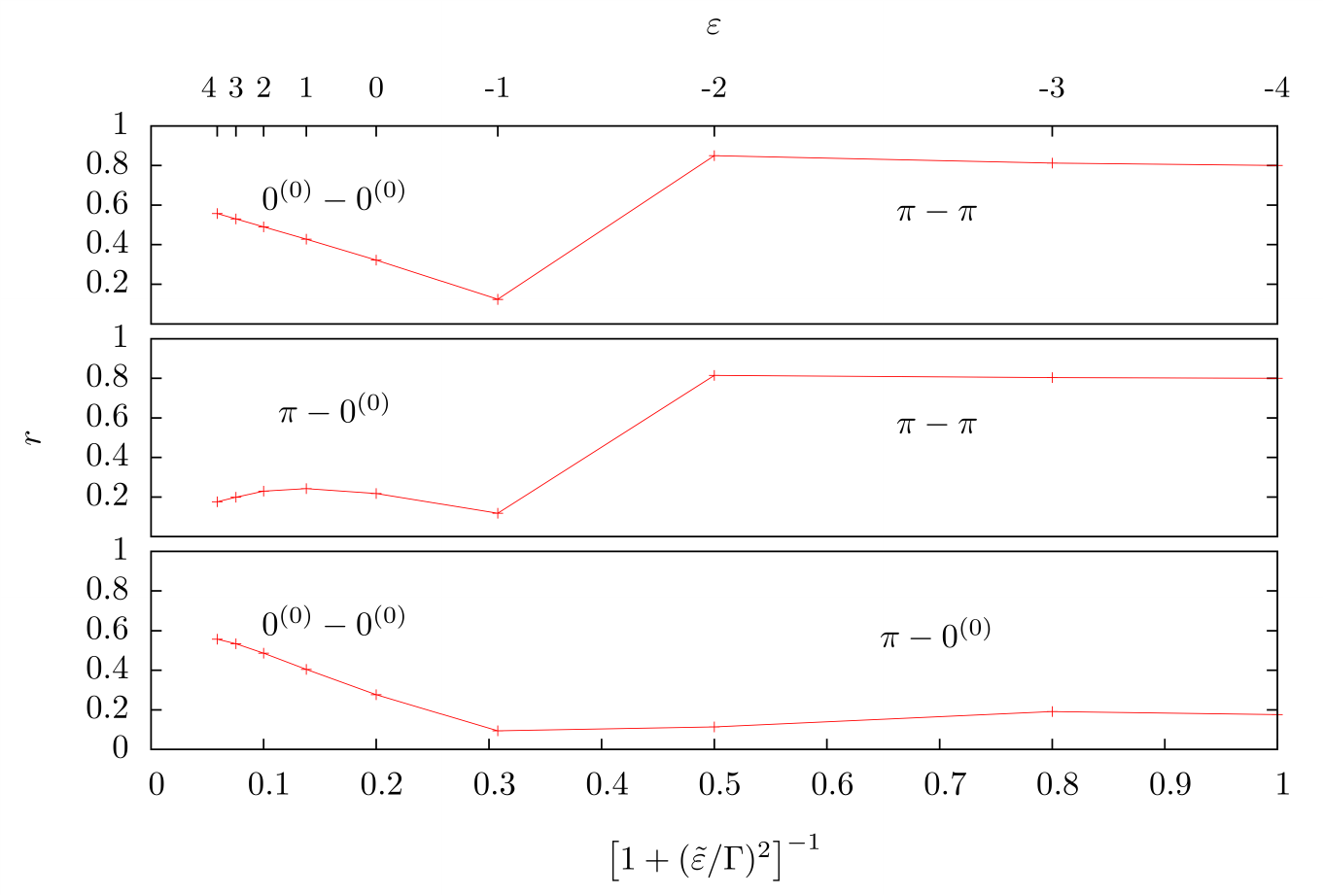}
\caption{Evolution of the splitting efficiency as a function of the effective transparency $D(\varepsilon)$ [Eq.~\eqref{trans_eff}], for fixed $\Gamma=2\Delta$ and $U=8\Delta$. 
\textit{Top:} The energy levels of both dots are taken to be identical and varied simultaneously ($\varepsilon_U=\varepsilon_D\equiv\varepsilon$). \textit{Middle:} the $D$ quantum dot is taken in the $\pi$ phase ($\varepsilon_D=-4\Delta$) while the energy of the $U$ dot is varied ($\varepsilon_U\equiv\varepsilon$). \textit{Bottom:} the $D$ quantum dot is taken in the $0^{(0)}$ phase ($\varepsilon_D=4\Delta$) while the energy of the $U$ dot is varied ($\varepsilon_U\equiv\varepsilon$). 
}\label{crossover_eff}
\end{figure}

Finally, we note that the energy levels of the dots can be easily varied experimentally using electrostatic gates. Varying the position of the energy level $\varepsilon$ of a quantum dot at fixed decay rate $\Gamma$ and fixed Coulomb on-site repulsion $U$ changes the effective transparency defined as
\begin{equation}
D(\varepsilon)=\frac{\Gamma^2}{\Gamma^2+\left(\varepsilon+\frac{U}{2}\right)^2}  .\label{trans_eff}
\end{equation} 
Thus, by varying the energy levels of the two dots, we can optimize the splitting efficiency by reaching the $\pi-\pi$ phase at high effective transparency. 
We show on the top panel of Fig.~\ref{crossover_eff} the splitting efficiency as a function of the energy 
level of the two dots, taken to be identical. The abrupt change of the efficiency near $D(\varepsilon) = 0.4$ shows the crossover from the $0^{(0)}-0^{(0)}$ to the $\pi-\pi$ junction. Similarly, when tuning only one of the two energy levels, while maintaining the other dot in the $\pi$ (middle panel) or the $0^{(0)}$ phase (bottom panel), the splitting efficiency shows a marked transition as a function of the effective transparency.
      
\section{Conclusion}\label{conclusion}

We have studied a double Josephson junction consisting of two quantum dots connected to two superconductors as a tool to probe Cooper pair splitting. When the two single Josephson junctions which constitute our device are coupled to each other via CAR processes, the doubling of the period of the critical current measured in a nanoSQUID experiment is an evidence for the emergence of nonlocal phenomena in the electronic transport of the double junction. This type of diagnosis may prove more convenient than non-equilibrium scenarios involving a superconducting source of electrons and two normal leads/dots where either nonlocal conductance signal or noise cross-correlations are measured. 

While this device had been studied in the context of perturbation theory in the tunneling Hamiltonian coupling the dot to the leads,\cite{choi_bruder_loss,wang_hu} no generalization to arbitrary transmission had been proposed, and no systematic study for optimizing the CAR signal with respect to the phases 
($0^{(0)}$, $\pi$, $0^{(2)}$) of the quantum dots had been attempted so far. The path integral approach of Ref. \onlinecite{rozkhov_arovas}, within reasonable approximations (saddle point treatment) allows precisely to meet these goals. One of the key results of the present work resides in defining the degree of efficiency of Cooper pair splitting, and evaluating it for the different dot configurations. 

We first studied the tunneling regime where the usual perturbative expression for the Josephson signal allows us to fit our numerical results and to introduce a natural definition of the splitting efficiency. We were also able to generalize this quantity to arbitrary transparency, providing a criterion for the efficiency of CAR processes which is based on an analysis of the AB signal of the Josephson critical current. We thus studied the prominence of nonlocal phenomena depending on the phases of the quantum dots and  found that the $\pi-\pi$ association optimizes the splitting of the Cooper pairs that are emitted from one superconductor and recombined on the other one. Yet, our analysis shows that the $0^{(0)}-0^{(0)}$ and $0^{(2)}-0^{(2)}$ combinations also provide robust Cooper pair splitting signals. Within each of these combinations of phases, we see for the most part that variations of the Coulomb repulsion parameter has little influence on the Cooper pair splitting efficiency. 

The present treatment should only be valid if the superconductor--dots system is above the Kondo regime.\cite{Kondo04,Siano04} For a single dot embedded between two superconductors, the Kondo effect manifests itself when the Kondo temperature is larger than the superconducting gap, because quasiparticle excitations are 
required to trigger the spin flip between the leads and the dot. The expression of the Kondo temperature for a Hubbard type Coulomb repulsion is given~\cite{Siano04,van_der_wiel} by 
$T_K =\sqrt{\Gamma U}/2\,\text{e}^{\pi\varepsilon_0 (\varepsilon_0 + U)/(\Gamma U)}$. Within the range of parameters chosen in our numerical study, we find that the Kondo temperature
is at most $0.13 \Delta$, which gives us confidence in our working assumptions. 
At any rate, the Kondo regime could be avoided by working with the $0^{(0)}-0^{(0)}$ combination of phases, which according to Fig.~\ref{eff_nonperturb} still has a sizable splitting efficiency. 

Furthermore, we treated the CAR coupling parameter $\eta$ in a phenomenological manner (varying it between $0$ and $1$), justifiably so because so far no experimental investigation of Cooper pair splitting in superconducting--normal metal ``forks'' out of equilibrium seems to find the power law suppression which is attributed to bulk 3D superconductors. This may be due to the fact that microscopic models have to be revisited, taking into account that electron emission/absorption in the superconductors is not point like, but should be averaged over some finite volume, reducing the effect of oscillations over the Fermi wavelength of the CAR parameter. 

Extensions of this work could be envisioned by going beyond the saddle point approximation of the Hubbard Stratonovich transformation.
This transformation is exact in its functional integral form and neglecting the fluctuations of the auxiliary field and then using a method of steepest descent are approximations which were sufficient to exhibit the $\pi$-shift in a single junction~\cite{rozkhov_arovas}. However, the possibility to go beyond this mean field type of approach could be considered by looking at Gaussian fluctuations around the stationary value of the auxiliary field. Alternatively, a numerical renormalization group method~\cite{Kondo04} could be employed in principle, but the proliferation of couplings and parameters is likely to render it cumbersome. 

\begin{figure}
\centering
\includegraphics[scale=0.35]{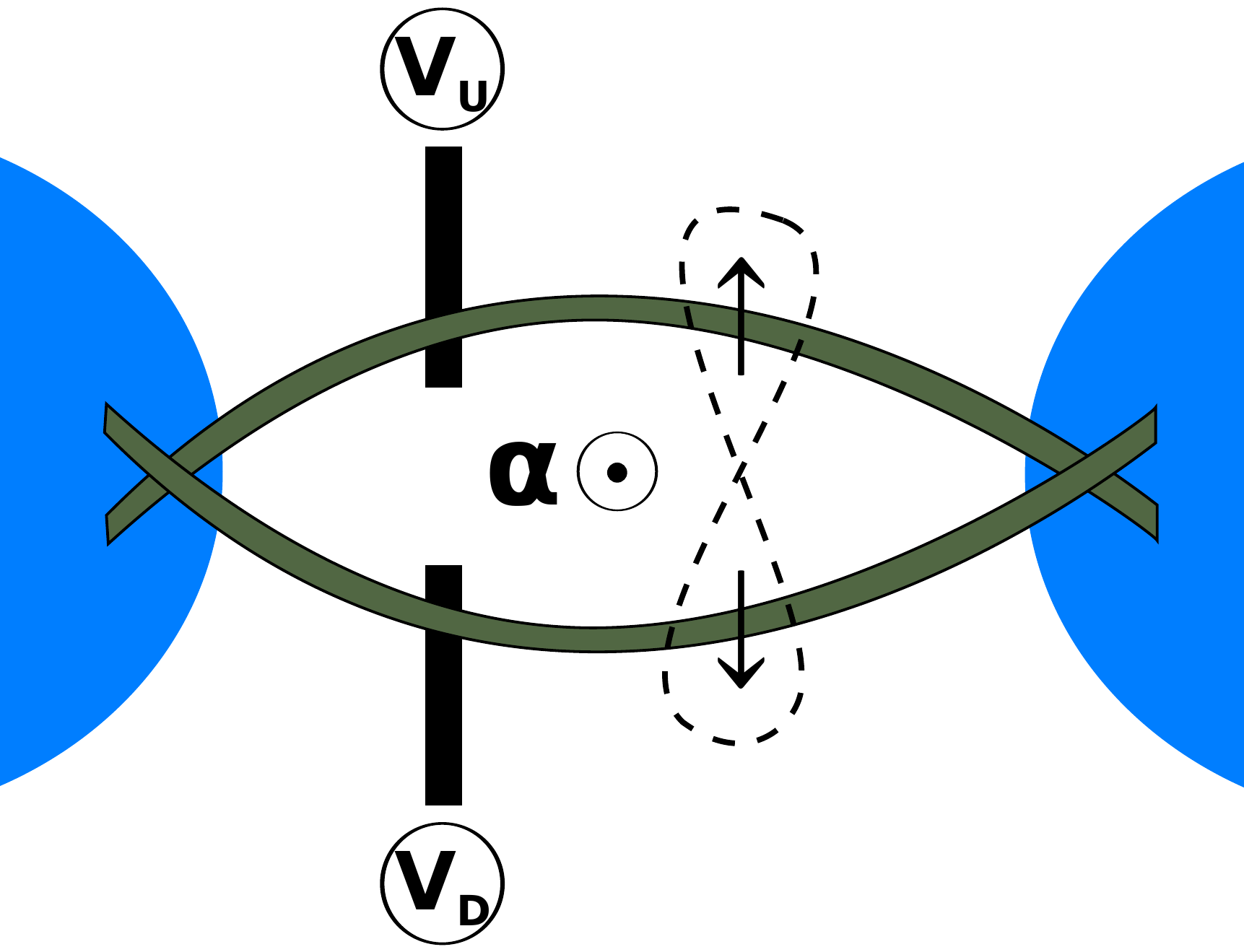}
\caption{Setup with nanowire/nanotube quantum dots.}\label{nanowire}
\end{figure}

A limitation of the present Cooper pair diagnosis resides in the fact that the nanoSQUID requires reduced dimensions so that the injection points to the dots on both superconductors are separated within a distance smaller than the coherence length. This imposes constraints on the separation of the quantum dots, and as a consequence, the area where the AB flux is imposed becomes reduced. Recall that in order to perform the AB diagnosis, a few flux quanta need to be introduced in this area, 
but if the imposed magnetic field needed to apply several flux quanta becomes larger than the critical field of the superconductors, the whole diagnosis breaks down. In order to optimize the surface area encompassed between the dots, we believe the best choice would be to work with nanowire/nanotube quantum dots, which have a large aspect ratio, and which nevertheless can achieve large charging energies even though their length may exceed several $\mu$m.\cite{tans} This possibility is illustrated in Fig.~\ref{nanowire}. To work out the numbers, we find that imposing 2 magnetic flux quanta within a 1 $\mu$m$^2$ area requires a magnetic field of $8\times10^{-3}$ Tesla, which is still smaller than the critical field of superconducting materials such as Aluminum ($10^{-2}$ Teslas) or Niobium (0.2 Teslas). Note that Aluminum could be quite adapted due to its long coherence length (1.6 $\mu$m). 

While this work came to completion, we became aware of a current biased measurement\cite{tarucha} which studies precisely the behavior of the double dot Josephson junction. There, a (remarkable) comparative study of the switching current (the current required to transit to the dissipative regime with voltage bias) was performed for different dot phase configurations. This experiment seems to provide realistic evidence of Cooper pair splitting, albeit indirectly, because the self organized quantum dots embedded in the Josephson junction are too close in order to impose the necessary magnetic flux to observe the AB oscillations.  

\acknowledgements
We acknowledge the support of the French National
Research Agency, through the project ANR NanoQuartets
(ANR-12-BS1000701). Part of this work
has been carried out in the framework of the Labex
Archim\`ede ANR-11-LABX-0033 and of the A*MIDEX
project ANR-11-IDEX-0001-02.
The authors acknowledge discussions with S. Tarucha and A. Oiwa about the nanoSQUID device. Discussions with E. Paladino and G. Falci are also gratefully acknowledged.

\bibliographystyle{unsrtnat}
\bibliography{biblio}

\end{document}